\documentclass[traditabstract]{aa} 
\usepackage{graphicx}
\usepackage{amssymb,amsmath}

\usepackage[varg]{txfonts}
\usepackage[pdftex=true, colorlinks=true, linkcolor=blue, filecolor=red,urlcolor=green]{hyperref}
\usepackage{natbib}
\def\Mp{M_{\rm p}}
\def\Rp{R_{\rm p}}
\def\Ms{M_{\star}}
\def\Msun{ M_\odot}

\def\Mearth{ M_\oplus}
\def\Rearth{ R_\oplus}

\usepackage{color}
\definecolor{green}{RGB}{0,200,0}

\def\figwidthmed{0.8\linewidth}
\begin{document}
   \title{The effect of rotation and tidal heating on the thermal lightcurves of \textit{Super Mercuries}} 

   \author{
   F. Selsis
          \inst{1,2} 
          \and
         A.-S. Maurin
         \inst{1,2}
         \and
         F. Hersant
         \inst{1,2}
         \and
         J. Leconte
         \inst{3}
         \and
         E. Bolmont
        \inst{1,2}
        \and
        S. N. Raymond
         \inst{1,2}
         \and
         M. Delbo'
         \inst{4}
   }

  \authorrunning{F. Selsis et al.}
 \titlerunning{Thermal lightcurves of Super Mercuries }

  \institute{
 Univ. Bordeaux, LAB, UMR 5804, F-33270, Floirac, France \\
 \email{selsis@obs.u-bordeaux1.fr,asophie.maurin@gmail.com,hersant@obs.u-bordeaux1.fr,rayray.sean@gmail.com} 
 \and
CNRS, LAB, UMR 5804, F-33270, Floirac, France
  \and
Laboratoire de M\'et\'eorologie Dynamique, Institut Pierre Simon Laplace, Paris, France \\
 \email{jeremy.leconte@lmd.jussieu.fr}
 \and
 UNS-CNRS-Observatoire de la C\^{o}te d'Azur, Laboratoire Lagrange, BP 4229, 06304 Nice Cedex 04, France \\
 \email{marco.delbo@oca.eu}
 }

  \date{accepted May 16, 2013}

 \abstract
{Short period ($<50$ days) low-mass ($<10\Mearth$) exoplanets are abundant and the few of them whose radius and mass have been measured already reveal a diversity in composition. Some of these exoplanets are found on eccentric orbits and are subjected to strong tides affecting their rotation and resulting in significant tidal heating.  Within this population, some planets are likely to be depleted in volatiles and have no atmosphere. We model the thermal emission of these \textit{Super Mercuries} to study the signatures of rotation and tidal dissipation on their infrared light curve. We compute the time-dependent temperature map at the surface and in the subsurface of the planet and the resulting disk-integrated emission spectrum received by a distant observer for any observation geometry. We calculate the illumination of the planetary surface for any Keplerian orbit and rotation.  We include the internal tidal heat flow, vertical heat diffusion in the subsurface and generate synthetic light curves. We show that the different rotation periods predicted by tidal models (spin-orbit resonances, pseudo-synchronization) produce different photometric signatures, which are observable provided that the thermal inertia of the surface is high, like that of solid or melted rocks (but not regolith). Tidal dissipation can also directly affect the light curves and make the inference of the rotation more difficult or easier depending on the existence of hot spots on the surface. Infrared light curve measurement with the James Webb Space Telscope and EChO can be used to infer exoplanets' rotation periods and dissipation rates and thus to test tidal models.  This data will also constrain the nature of the (sub)surface by constraining the thermal inertia. }


   \maketitle
%

\section{Introduction}
   \begin{figure}[b]
	\centering
	\includegraphics[width=\figwidthmed]{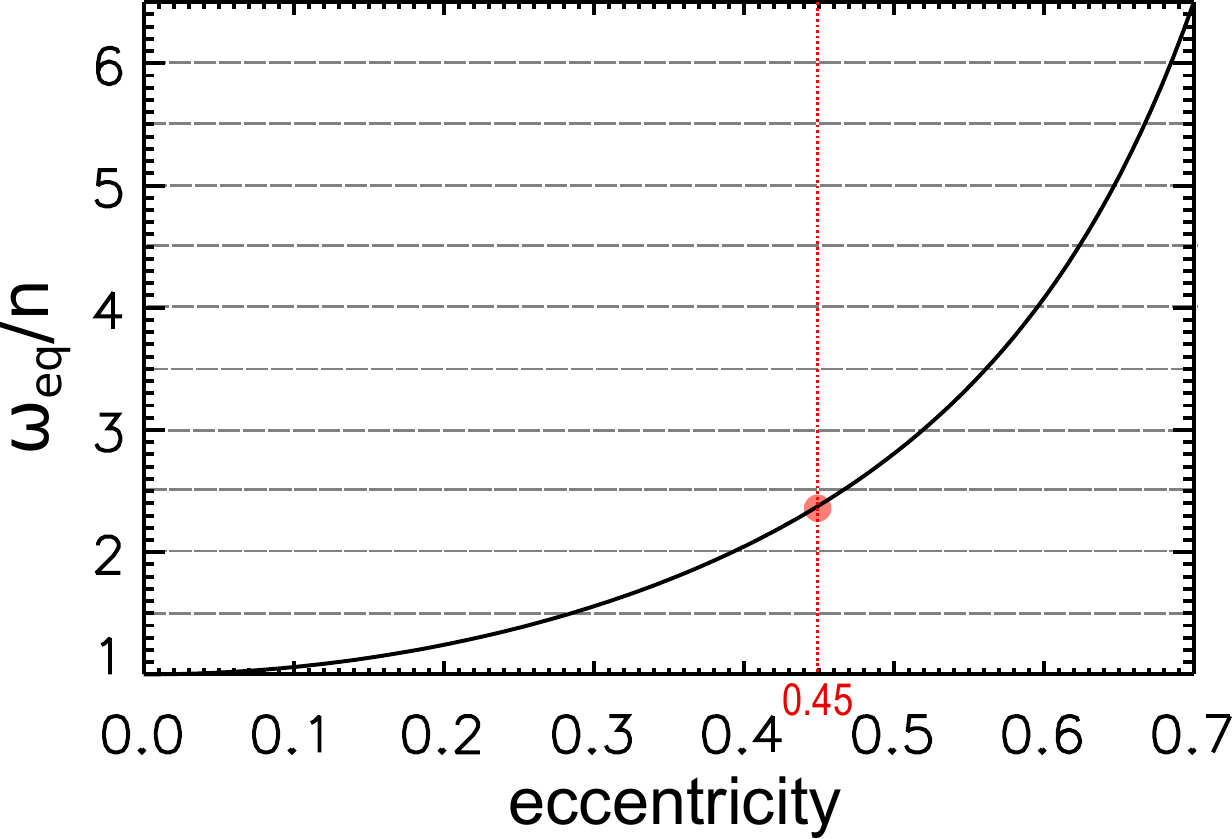}
 	\caption{Number of rotations per orbital period for a pseudo-synchronized planet as a function of eccentricity. The filled circle shows the pseudo-synchronization ($\omega_{\rm eq}/n=2.38$) for the default value of eccentricity used in this study ($e=0.45$).} 
	\label{fig:equ_rot}%
\end{figure}
Exoplanet Doppler searches have revealed a high frequency of exoplanets in the $1-10\Mearth$ mass range. Around F, G, K stars, \citet{Mayor2011} found a debiased occurrence rate $f=0.41\pm0.16$ for these masses and orbital periods shorter than 50~days. More than one quarter of all F-G-K stars host this type of \textit{hot Super-Earth}, as they are sometimes called. For the same types of stars, orbital periods, and the $1-1.8\Rearth$ range (corresponding to $1-10\Mearth$ for rocky planets), \citet{Petigura2013} found an occurrence $f\sim17\%$ from Kepler candidates, a result that significantly departs from the estimates derived from radial velocity detections. From the HARPS survey of M stars and for the same mass range, \citet{Bonfils2013} inferred a frequency of $0.36^{+0.25}_{-0.10}$ and $0.50^{+0.52}_{-0.16}$ for orbital period within $1-10$~d and $10-100$~d, respectively.  Statistics of Kepler candidates for cool stars imply a similar frequency: 0.51 planet per star for the $0.5-1.4\Rearth$ range and periods shorter than 50~days \citep{Dressing2013}. \\

Despite the abundance of Kepler transiting candidates with radii below $2 \Rearth$, only a few planets have well-measured masses below $10\Mearth$.  These low-mass planets exhibit a variety of densities that goes from lower than 1~g/cm$^3$ for Kepler~11~c,e,f \citep{Lissauer2013} up to $8.8\pm2.9$~g/cm$^3$ for Kepler~10~b \citep{Batalha2011}. Within this mass range, we can expect to find volatile-rich planets with dense atmospheres but also airless bodies that either accreted from dry, refractory material or lost their volatile component due to the strong irradiation. At  0.04~AU from a Sun-like star, a $5\Mearth$ planet intially containing 10\% of its mass in H$_2$O (a proportion 100 times higher than that of Earth) could lose its whole volatile content in 500~Myrs \citep{Selsis_OP_2007}. \\

\begin{table*}
\centering
\begin{tabular}{lccccl}
\hline
planet & $\Ms (\Msun)$ & $M\sin i(\Mearth)$& Period (d) & eccentricity & reference \\
\hline
 HD~181433~b & 0.78  & 7.5& 9.37 & $0.396 \pm 0.062$ & \citet{Bouchy2009} \\
 GJ667C~c & 0.33 & 4.25 & 28.1 & $0.34 \pm 0.1$ & \citet{Delfosse2012} \\ 
 GJ581~e & 0.31 & 95 & 3.14 &  $0.32 \pm 0.09$ & \citet{Forveille2011} \\
 GJ581~d & 0.31 & 6.1 & 66.6 &  $0.25 \pm 0.09$ & \citet{Forveille2011} \\
 GJ876~d & 0.3  & 6.83 & 1.93 & $0.257 \pm 0.07$ & \citet{Rivera2010} \\
 GJ674~b & 0.3 & 11.1 & 4.7 & $0.20\pm 0.02$ & \citet{Bonfils2007} \\
HIP~57274~b & 0.73 &11.6  & 8.13 & $0.19\pm0.1$ & \citet{Fischer2012} \\
$\mu$~Arae~c & 1.08 & 10.5 & 9.63 & $0.172\pm 0.040$ & \citet{Pepe2007} \\
\hline
\end{tabular}
\caption{Some known low-mass exoplanets on eccentric orbits.}
\label{table:planets}
\end{table*}

Although most of short-period low-mass exoplanets have circular orbits, some have significant eccentricities. Their orbit could have been excited by other planets in the system, continuously or at some moment in the past, or circularization might be slow due to a low dissipation. Table~\ref{table:planets} gives a list of known exoplanets found by radial velocity surveys with $4.5 < m\sin i < 12\Mearth$ and $0.1<e<0.45$.\\

For planets on eccentric orbits and subjected to strong tides the rotation period is not known a priori. The planet can be captured into a spin-orbit resonance like Mercury, which rotates three times every two orbits (resonance 3:2). It can also have reached pseudo-synchronization, a theoretical state that minimizes the dissipation \citep{Hut1981, Levrard2007}. This equilibrium rotation ($\omega_{\rm equ}/n$, see formula~\ref{equ:equ_rot} and Fig.~\ref{fig:equ_rot}) is always faster than synchronization (1:1 resonance).
The existence of the pseudo-synchronous state has recently been put in question for the case of solid planets by \citet{Makarov2013}, who claim that only spin-orbit resonances should be expected. The planet can, however, be captured into different spin-orbit resonances depending on its initial rotation, the evolution of its eccentricity, its internal structure and rheology, its oblateness (which depends on its rotation rate) and interactions with other planets. In general, the order of the most likely spin-orbit resonances is expected to increase with the eccentricity \citep{MBE12}. Measuring the rotation period of eccentric exoplanets would therefore provide valuable information for tidal theories as well as useful constraints on the nature and history of the planets. In the present study, we explore how rotation can be measured or constrained thanks to the infrared photometric variations of the planet.   \\

In addition to transits and eclipses when they occur, the photometric variation of a spatially-unresolved planetary system contains two superimposed signals: the intrinsic variability of the star, and the change in apparent brightness of the planet(s). One obvious planetary modulation is associated with the orbital phase. This phase modulation (or phase curve), which is obvious in reflected light, also exists at thermal wavelengths provided that a sufficient brightness temperature contrast exists between the day and night sides at the wavelength of the observation. Phase curves have been observed for hot jupiters both in the infrared \citep{Knutson2007} and in visible light \citep{Welsh2010}, for both transiting and non-transiting planets \citep{Crossfield2010}, and for eccentric planets \citep{Lewis2013,2013arXiv1303.5468C}. The reflected light curve of Kepler~10~b, a very short-period transiting "terrestrial" planet, was obtained by \citet{Batalha2011} despite a relative amplitude of only $\sim5\times10^{-6}$ (5~ppm), and the day-side emission of a similar exoplanet, 55~Cnc~e was measured with Spitzer \citep{Demory2012}. \\
Thermal and reflected light curves have been modeled for multi-planet systems \citep{Kane_55Cnc_2011,Kane_multiplanets_2013}. Atmosphere models have been used to compute multiband phase curves, assuming radiative equilibrium \citep{Barman2005}, a time-dependent radiative-convective model \citep{Iro2005} and more recently 3D general circulation models \citep{Showman2009,Selsis2011,Menou2012}. The photometric signature of eccentric exoplanets has been studied by several authors \citep{Iro2010, Kane_Gelino_periastron_2011,Lewis2013}. \citet{Maurin2012} modeled the spectral phase curves of synchronously rotating \textit{Super Mercuries} and showed that the inclination, albedo and radius of the planet could be inferred for out-of-transit configurations by multi-wavelength observations. Lightcurves of multi-planet systems have been modeled by \citet{Kane_55Cnc_2011} 
and \citet{Kane_multiplanets_2013}. \citet{Bolmont_55Cnc} investigated the influence of tidal heating on the thermal emission and phase curve of 55~Cnc~e. 

Thermal light curves have been proposed as a technique to characterize habitable Earth-like planets and to detect the presence of a dense atmosphere with future instruments \citep{2004ASPC..321..170S}.  Several aspects of characterizing habitable worlds using orbital photometry have been studied: the effect of the planet's obliquity \citep{2004NewA...10...67G} and eccentricity \citep{Cowan2012}, the presence of a moon \citep{2004ASPC..321..170S,2009AsBio...9..269M}, and the rotation period in the visible \citep{ford2001,2008ApJ...676.1319P} and infrared \citep{Illeana2012}. \\

Observing thermal phase curves of planets smaller than $2\Rearth$ will require future space observatories like JWST and EChO \citep{EChO2012}. Today, it may be achievable with Spitzer only for a very hot and very nearby object like 55~Cnc~e. EChO aims at performing high precision spectro-photometry with an ability to detect relative photometric variations better than 10~ppm with days to weeks stability and a broad spectral coverage ($0.4-16\mu$m). Observing phase curves to characterize low-mass exoplanets and their atmosphere is an important part of the science of EChO.\\


Here we study the photometric signature of \textit{Super-Mercuries}, big rocky exoplanets with no atmospheres on eccentric orbits. In section~\ref{sec:model}, we describe the insolation, thermophysical and tidal models used to produce thermal light curves. In section~\ref{sec:results}, we present the results and describe the influence of rotation, eccentricity, observing geometry,  thermal inertia  and tidal dissipation on the light curves. Our approach and results are discussed in section~\ref{sec:discussion} and the main conclusions of the study are given in section~\ref{sec:conclusion}.

\section{Model} \label{sec:model}
\subsection{Illumination of the planetary surface}
Our first step is to compute the time-dependent orbital distance $r$ and zenith angle $\theta$ at any point of the planetary surface. We assume a Keplerian orbit and a rotation defined by a period, obliquity and argument of periastron. Although our model can be applied to any orientation of the rotation axis, we do not consider oblique planets in this study and the (prograde) rotation is thus only defined by its period. 

The bolometric stellar flux $\phi_{\star}$ absorbed by the planetary surface at a given longitude and latitude and at an orbital distance $r$ is given by 
\begin{equation}
\phi_{\star}=\epsilon \sigma T_{eff}^4 \left(\frac{R_{\star}}{r}\right)^2 (1-A) \mu
\end{equation}
where $\sigma$ is the Stefan-Boltzmann constant, $T_{eff}$ and $R_{\star}$ are the effective temperature and radius of the star, $A$ is the surface bolometric albedo (assumed to be constant over the planet and thus equal to the Bond albedo), $\mu$ the cosine of the zenith angle $\theta$ and $\epsilon$ the bolometric emissivity.  Both $r$ and $\mu$ are time-dependent. The distance $r$ is obtained by solving Kepler's equation by Newton-Raphson iterations using the cubic approximation by Mikkola \citeyearpar{1987CeMec..40..329M} as a first guess (IDL keplereq routine). For a null obliquity the expression for $\mu$ is given by
\begin{eqnarray}
\mu &= &\cos(\theta)= \cos(\textrm{lat}) \cos(\textrm{lon}-\textrm{lon}_{\star}(t)), \, \textrm{if} \, \cos(\theta) \geq 0, \\
\mu&=&0, \, \textrm{if} \, \cos(\theta) < 0, \nonumber
\end{eqnarray}

 where $\textrm{lon}_{\star}(t)$ is the longitude of the substellar point, which can be calculated as follows for a null obliquity and a prograde rotation:
\begin{equation}
 \textrm{lon}_{\star}(t) = \nu(t) - \omega t + C 
 \end{equation}
 where $\nu$ is the true anomaly of the planet, $\omega$ the angular spin velocity and $C$ is an arbitrary constant depending on the position of the planet at $t=0$ and the origin of the longitude coordinate. If $\nu(t=0)$ and $C$ are set to $0^{\circ}$ then the planet is initially at periastron with the substellar point at $0^{\circ}$ longitude.

\subsection{Internal heat flow from tidal dissipation} \label{sec:tidal_model}
In this study, we neglect the internal heat flux due to the decay of radiogenic species or the release of the energy of accretion/differentiation, which should only affect the planetary thermal emission at a measurable level for very young objects. We include the surface heat flow resulting from tides given by $\Phi_{i} = \dot{E}_{tides}/4\pi R_{P}^2$ where $\dot{E}_{tides}$ is the rate of tidal dissipation. To calculate the dissipation rate, we use the constant time-lag model described in Leconte et al. \citeyearpar{Leconte2010}. For a planet with a mass $\Mp$ orbiting a star with mass $\Ms$ at a semi-major axis $a$ and with a null obliquity, the rate of tidal dissipation is given by: 
\begin{equation}
\dot{E}_{tide} = 2 K_p \left[ N_a(e) -2N(e)\frac{\omega}{n} + \Omega(e) \left( \frac{\omega}{n} \right)^2 \right],
\label{equ:dissip}
\end{equation}
with
\begin{align}
N(e) &= \frac{1+\frac{15}{2}e^2+\frac{45}{8}e^4+\frac{5}{16}e^6}{(1-e^{2})^{6}},\\
N_a(e) &= \frac{1+\frac{31}{2}e^2+\frac{255}{8}e^4 + \frac{185}{16}e^6 + \frac{25}{64}e^8}{(1-e^{2})^\frac{15}{2}},\\
\Omega(e) &= \frac{1+3e^2+\frac{3}{8}e^4}{(1-e^{2})^\frac{9}{2}}.
\end{align}
where $e$ is the eccentricity, $n$ is the orbital mean motion, $\omega$ is the rotation rate of the planet.  The quantity $K_{\rm p}$ is defined as:
\begin{equation}
K_{\rm p} = \frac{3}{2} k_{2,{\rm p}} \Delta t_{\rm p} \left( \frac{G\Mp^2}{\Rp}\right) \left( \frac{\Ms}{\Mp}\right)^2
\left( \frac{\Rp}{a}\right)^6 n^2
 \end{equation}

where  $G$ is the gravitational constant, $\Rp$ and $\Mp$ are the radius and mass of the planet, and $\Ms$ is the mass of the star. $\Delta t_{\rm p}$ and $k_{2,{\rm p}}$ are the time-lag and potential Love number of degree 2 for the planet, for which we use the Earth quantities ($\Delta t_{\rm p}=630$~s and $k_{2,{\rm p}}=0.3$) as the default dissipation \citep{DeSurgyLaskar1997}.

In this study, we assume that eccentric planets are either captured into a spin-orbit resonance or have reached the equilibrium rotation rate (or pseudo-synchronization) that minimizes $ \dot{E}_{tides}$. This equilibrium rotation rate is a function of $e$ (see Fig.~\ref{fig:equ_rot}):
\begin{equation}
\omega_{\rm eq} = \frac{N(e)}{\Omega(e)} n  \label{equ:equ_rot}
\end{equation}
\begin{figure*}
	\centering
	\includegraphics[width=\figwidthmed]{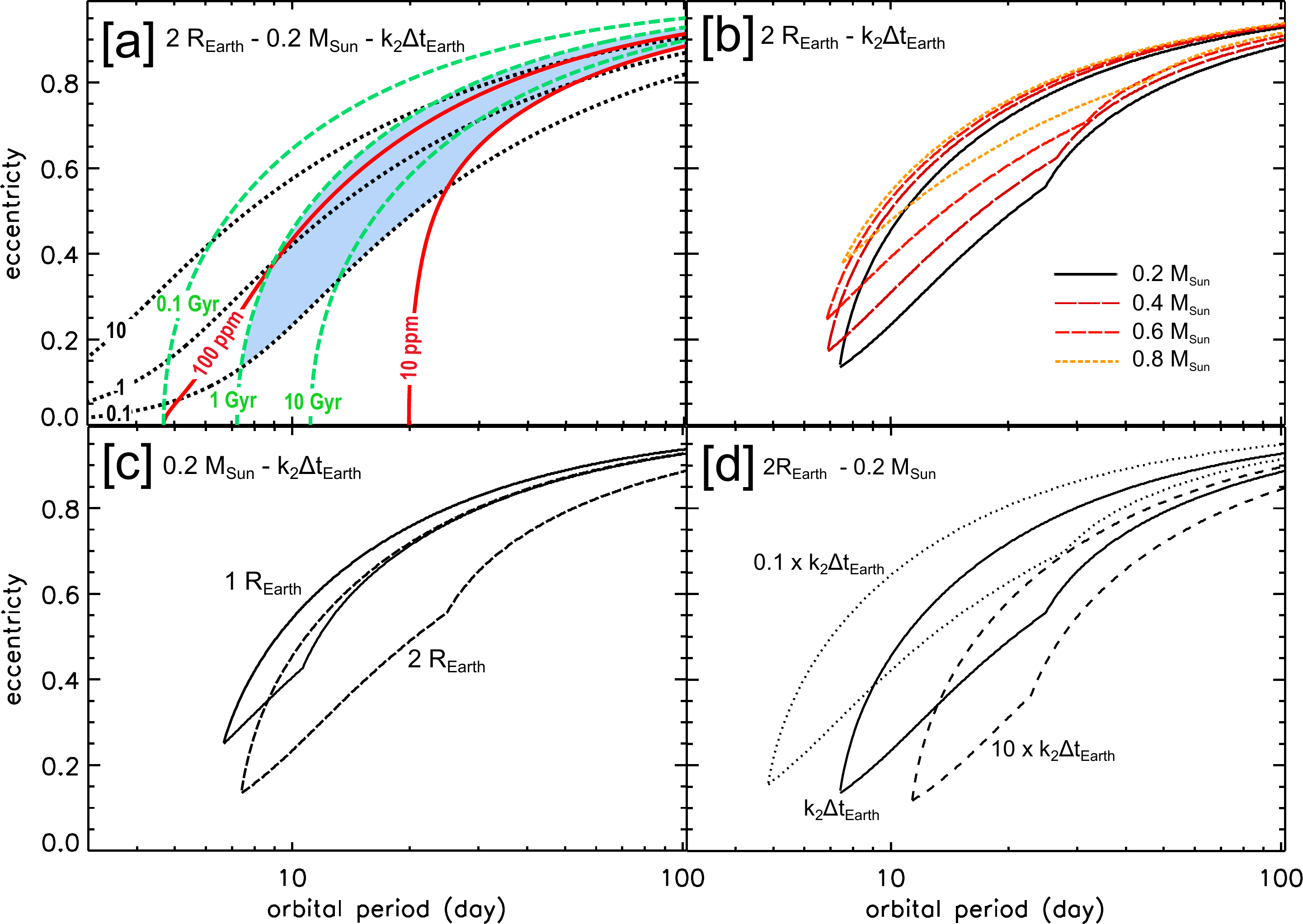}
 	\caption{Observability of the 10~$\mu$m emission excess due to tidal heating in a period-eccentricity diagram. a) Dashed contours indicate the circularization timescale ($e/\dot{e}$), solid contours give the planet/star contrast ratio in ppm, and the dotted contours show the ratio between the emission due to tidal heating and the emission without tidal heating. The contours are calculated for a $0.2\Msun$ star, $A=0.2$, an isothermal surface (stellar and internal flux fully redistributed), a $2\Rearth$ planet and the dissipation parameters of the Earth. The colored area represents the observability domain for the tidal heating, where the configuration lasts for more than 1Gyr, produce a planet/star contrast higher than 10~ppm and results in tidal heating contributing to more than 10$\%$ of the emission.  Panels b, c and d respectively show the influence of the stellar mass, the planetary radius and the dissipation factor on this tidal heating observability domain. }
	\label{fig:tides_param}%
\end{figure*}
Figure~\ref{fig:tides_param} shows the eccentricity-orbital period domain in which three observation criteria are fulfilled. First, tidal dissipation increases the planet thermal emission by at least 10\%. Second, the flux from the planet exceeds $10^{-5}$ times that of the star at $10~\mu$m.  Third, the eccentricity damping time $e/\dot{e}$ is longer than 1~Gyr. This damping timescale is calculated according to \citet[][eqn 6]{Leconte2010} assuming no forcing from planetary companions. Note that perturbations from massive or eccentric planets in the same system could maintain a high eccentricity for longer~\cite{Beust2012,Bolmont_55Cnc}. This criterion is therefore conservative and the observability domain could be extended to higher eccentricities in multiple systems. Fig.~\ref{fig:tides_param} also shows how this observability domain varies with the mass of the central star, the radius of the planet and the dissipation factor (which is poorly constrained for exoplanets). In these graphs the thermal emission of the planet is calculated assuming an isothermal surface heated by the fully redistributed stellar illumination (averaged over one orbital period) and the tidally-produced internal heat flow.

\subsection{Subsurface heat diffusion}
We compute the surface and subsurface time-dependent temperature taking into account the internal heat flow and the forcing by the absorbed stellar light by modeling the vertical diffusion of heat. 

The 1D heat equation can be written: 
\begin{eqnarray}
\rho \ c(T) \frac{\partial T(z,t)}{\partial t} & = & \frac{\partial}{\partial z} \left [\kappa(T) \frac{\partial T(z,t)}{\partial z} \right ]
\label{diffeq}
\end{eqnarray}
with $z$ the depth, $\rho$ the density of the layer, $\kappa$ the thermal conductivity and $c$ the heat capacity at constant volume (usually noted $c_V$).  \citet{MaurinThese} investigated the effect of using a realistic temperature-dependency for $c(T)$ and $\kappa(T)$ compared with using constant values for $c$ and $\kappa$ calculated for the planet's equilibrium temperature $T_{\rm eq}$.   \citet{MaurinThese} conducted a broad exploration of the parameter space of $T_{\rm eq}$, rotation rate, and surface materials, and found maximum differences of only a few K locally in the surface temperature with a negligible impact on the disk-integrated thermal emission. Therefore, and although our model can include the temperature dependency of $\kappa$ and $c$, we assume $\kappa$ and $c$ to be constant. As in most heat diffusion models used in planetary science~\citep{Spencer1989}, Eq. (\ref{diffeq}) simplifies to:
\begin{eqnarray}
\rho c\frac{\partial T}{\partial t} =\kappa\left(\frac{\partial^2T}{\partial z^2}\right)
\label{eqdiffsimple}
\end{eqnarray}

\subsubsection{Boundary conditions}
At a given longitude-latitude point of the planet, the upper-boundary conditions is determined by the absorbed stellar light and the thermal emission is given by
\begin{eqnarray}
\kappa \left(\frac{\partial T(z,t)}{\partial z}\right)_{z=z_{max}} = \epsilon \sigma T^4-\phi_{\star}(\textrm{t})
\label{condlim}
 \end{eqnarray}
 where $z_{max}$ corresponds to the surface and $\epsilon$ is the surface emissivity (assumed to be equal to 1 in this study).
  The lower-boundary condition is given by the internal heat flow:
 \begin{eqnarray}
\Phi_{i}&=&\kappa \left(\frac{\partial T(z,t)}{\partial z}\right)_{z=z_{min}}
\label{fluxint}
 \end{eqnarray}
where $z_{min}$ is the bottom of the modeled vertical column and $\Phi_{i}$ is computed from the tidal dissipation. We assume that $z_{min}$ is large enough that the variations of surface irradiation do not propagate down to $z_{min}$ such that temperature eventually reaches a steady-state.
\subsubsection{Dimensionless equations}
We introduce a timescale $\tau$ (without initially specifying it) and the typical length over which diffusion acts in a duration $\tau$
\begin{eqnarray}
 L & = &  \sqrt{\frac {\kappa \tau}{\rho c}}.
\end{eqnarray}
These quantities can then be used to make the change of variables
\begin{eqnarray}
\tilde{t} =  \frac{t}{\tau}\label{deftau} & & \tilde{z} = \frac{z}{L}\label{defz}.
\end{eqnarray}
In the dimensionless variables, Eq. \ref{eqdiffsimple} takes the simple form:
\begin{eqnarray}
\frac{\partial T(\tilde{z},\tilde{t})}{\partial \tilde{t}} =\frac{\partial^2T(\tilde{z},\tilde{t})}{\partial \tilde{z}^2},
\label{eqdiffsimple3}
\end{eqnarray}
while equations \ref{condlim} and \ref{fluxint} become
\begin{eqnarray}
\label{condlim2}
\sqrt{\frac{1}{\tau}} \sqrt{\kappa \rho c} \left(\frac{\partial T(\tilde{z},\tilde{t})}{\partial \tilde{z}}\right)_{z=z_{max}}& =& \epsilon \sigma T^4-(1-A)\phi_\star(t)\\
\sqrt{\frac{1}{\tau}} \sqrt{\kappa \rho c} \left(\frac{\partial T(\tilde{z},\tilde{t})}{\partial \tilde{z}}\right)_{z=z_{min}}& =& \Phi_{i}.
 \end{eqnarray}
Once the illumination flux $\phi_\star(t)$ and the internal flux $\Phi_{i}$ are known, 
the surface temperature $T({z},{t})$ is determined by a single physical parameter, the thermal inertia $\Gamma$:
\begin{eqnarray}
\Gamma& = & \sqrt{\kappa\rho c}
\label{TIeq}
\end{eqnarray}
\subsubsection{Numerical scheme}

To calculate the temperature structure we discretize Eq. \ref{diffeq} assuming layers of constant thickness.  To ensure conservation, the temperatures are defined at the center of each layer and fluxes are calculated at each interface.
The system is then solved using a Crank-Nicolson
scheme, with Neumann boundary conditions computed from eqs. \ref{condlim}
and \ref{fluxint}. 
One version of our model uses a variable thickness (small near the surface, larger close to the lower boundary) without significantly reducing the required number of layers.  

In order to set the thickness of the layers and the total depth of the column, we need to evaluate the characteristic length scales of the problem. If the surface insolation is periodic with a pulsation $\omega_\star$, we can use $\tau = 1/\omega_\star$ and $L=l_s$, where $l_s$ is the thermal skin depth as defined by \citet{Spencer1989}:\\
\begin{eqnarray}
l_s & = & \sqrt{\frac{\kappa}{\rho c \omega_\star}}
\label{TSeq}
\end{eqnarray}
For an eccentric orbit, the insolation at a given point of the surface is not periodic.  The length of the day -- or duration between two consecutive star rises -- varies throughout the orbit, with the exception of some spin-orbit resonances. In addition, the zenith angle does not vary periodically (except, again, for some spin-orbit resonances). We consider two timescales: the mean diurnal cycle and the orbital period, and their associated length. The illumination modulation with the shortest periodicity determines the value of the thermal skin and imposes the vertical spatial resolution.  If the orbital motion induces the shortest periodicity then the thermal skin $l_s$ is calculated assuming $\omega_\star= n $, where $n$ is the mean motion.  If the averaged length of the day (calculated assuming a mean orbital motion) is shorter than the orbital period then the thermal skin $l_s$ is calculated using $\omega_\star=\omega_{\rm rot} - n$, where $\omega_{\rm rot}$ is the pulsation associated with the sidereal rotation. The modulation with the longest period sets the depth of the column to be modeled. In practice, however, we use only one single (the smallest) value for the thermal skin $l_s$ and typically place our deepest layer at 10 to $20l_s$, using 64 to 256 layers, depending on the case. 

This method has the advantage of simplicity (compared for instance with a time-dependent adaptative-grid) and effectiveness. However, one should keep in mind that using a resolution based on an average value for $\omega_\star$ and $l_s$ may prevent the model from resolving heat waves with very short wavelengths that are produced during a small fraction of the orbit, for instance near periastron for highly eccentric orbits. This is one reason why we systematically verify that the spatial resolution and total depth we use are sufficient, by making sure that the results remain unchanged after an increase of the number of layers and/or the total depth. 

Temperature profiles are typically calculated for 684 surface points (19 latitudes, 36 longitudes), enough to produce smooth light curves. A smaller number of points is used when testing the sensitivity to initial conditions, the required total depth, number of layers, and number of orbits. 
As the use of the model is not limited by computation time we can afford to use a small and constant time step (typically 0.2\% of the orbital period), which is the same for all points on the planetary surface. \\

\begin{table*}
\centering
\begin{tabular}{cccccccccccc}
\hline
$\Ms$ & orbital & $e$ & $\Rp$ &$ \Gamma$ & $\kappa$ & $A$ & $\Delta t_{\rm p}$ & $k_{2,{\rm p}}$ & inclination & longitude of   & $\lambda$ \\
($\Msun$)& period (d)&  & ($\Rearth$) & Js$^{-1/2}$K$^{-1}$m$^{-2}$ & Wm$^{-1}$K$^{-1}$ &  & (s) & &($^{\circ}$) &   the observer  & ($\mu$m) \\
 \hline
0.3 & 10 & $0.45$ &2 & 3000 & 3 & 0.1 & 630 & 0.3 & $90$ &$2^{(*)}$ & $10$ \\
 \hline
\end{tabular}
\caption{Default parameters of the study (unless other values are specified). $^{(*)}$ refers to Fig.~\ref{fig:long_var}.}
\label{table:default}
\end{table*}

\subsubsection{Initial temperature profile}
For the result to be independent of the arbitrary initial (sub)surface temperature profile, a minimum number of orbits must be simulated. 
To keep this number as small as possible it is important to start from initial conditions that are as close as possible to an average thermal state. For each point on the planet, we calculate $\bar{\phi_\star}$, the mean stellar energy absorbed over all orbits of the simulation. In the absence of an internal heat flux we start with an isothermal profile with temperature given by $T=(\bar{\phi_\star}/\epsilon \sigma)^{1/4}$. If there is an internal heat flux $\Phi_{i}$ then the initial temperature profile is:
\begin{eqnarray}
T_{\rm surf} &=&\left(\frac{\bar{\phi_\star}+\Phi_{i}}{\epsilon \sigma} \right) ^{\frac{1}{4}} \\
\frac{\partial T(z)}{\partial z} &= &\frac{\Phi_{i}}{\kappa}
\end{eqnarray}
We note that the real time-averaged temperature and internal energy differ from this initial profile, even in the simplest case of radiative equilibrium and no internal heat flux. This is a consequence of the Rogers-H\"olders inequality:
$$
\frac{1}{\Delta t} \int_{0}^{\Delta t} T(t) \,\mathrm{d}t \leq  \left( \frac{1}{\Delta t} \int_{0}^{\Delta t} T^4(t)\,\mathrm{d}t \right)^\frac{1}{4}
$$
Because we calculate a temperature from an averaged flux, this initial condition is always too hot but the heat excess is not known a priori. For large variations of $T(t)$ this excess can be significant, especially if $T(t)$ approaches 0~K. For this reason, it is important to check that simulations have been run over a sufficient number of orbits for the modeled subsurface to release this excess heat. Note that the deeper the modeled column and the larger the thermal inertia the longer it takes to transport the initial excess heat to the surface and radiate it to space. The total depth and thermal inertia thus have an important effect on the required number of orbits. In most of our simulations 12 orbits were sufficient but some cases required up to 36 orbits. This number also depends on the actual duration of an orbit.
\begin{figure}
	\centering
	\includegraphics[width=\figwidthmed]{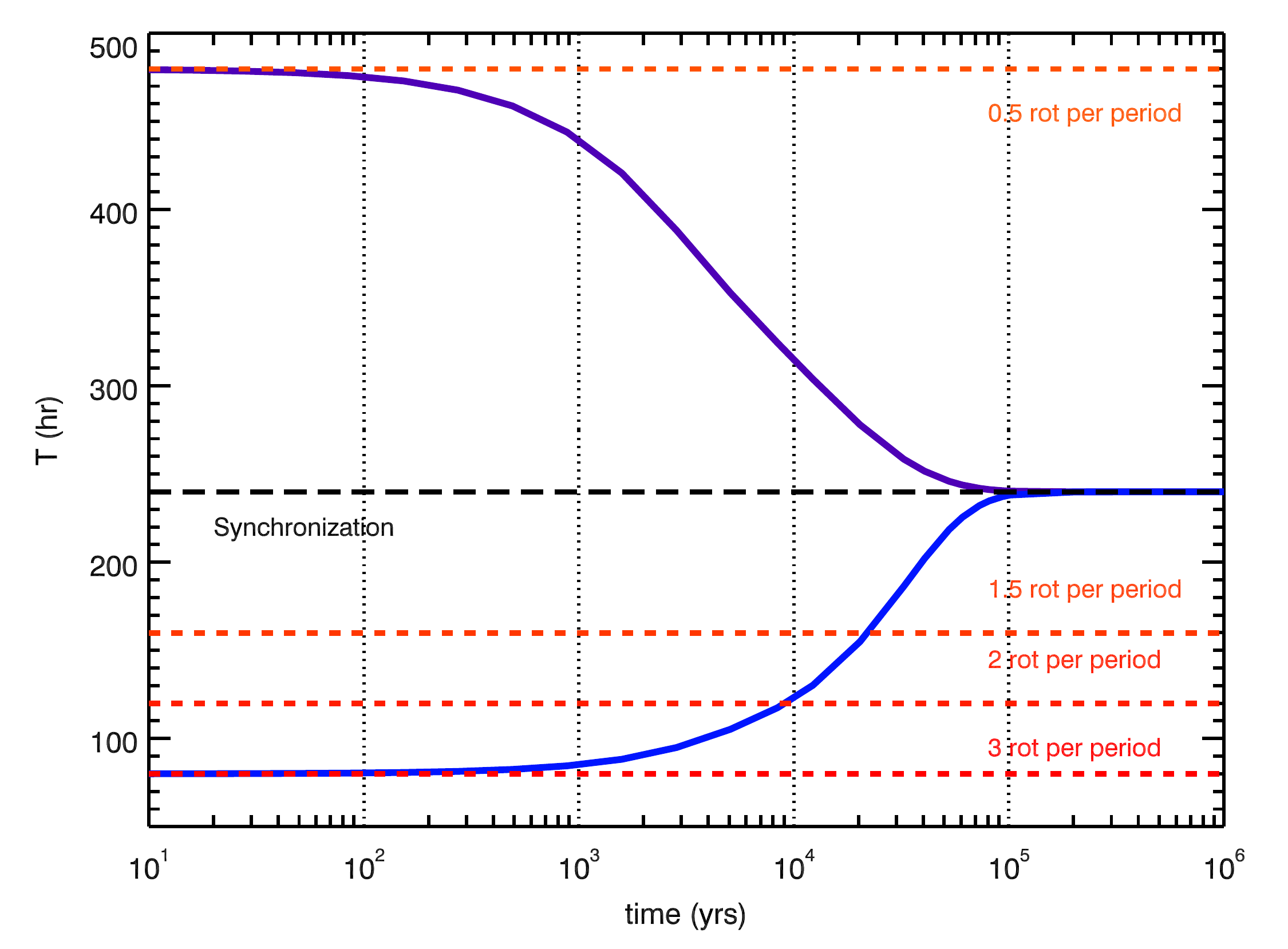}
 	\caption{Evolution of the rotation period of a planet starting from the 3:1 and 1:2 spin-orbit resonances. All the parameters of the system, except $e=0$, are set to the default values of table~\ref{table:default}. }
	\label{fig:synchro}%
	\end{figure}
\subsection{Lightcurves for a distant observer}
For the configurations and at the wavelengths considered in this work, the contribution from reflected light remains negligible compared with the thermal emission. We do not include the contribution of the reflected light in the formula, although it is calculated by the model.
The surface of the planet is divided into a longitude-latitude grid. Each cell $j$ of the grid has a surface temperature $T_j$ and an area $S_j$. The flux spectral density received by a distant observer at a distance $d$ is 
$$ \phi_{\rm p,\lambda}(d) = \sum_j I_{\lambda,j} \frac{S_j\cos\alpha_j}{d^2}$$
where $\alpha_j$ is the angle between the normal to the cell and the direction toward the observer and $I_j$ is the specific intensity of the cell given by
$$I_{\lambda,j} = \frac{\epsilon_{\lambda} B_{\lambda}(T_j)}{\pi}$$
where $B_{\lambda}$ is the Planck function, $\epsilon_\lambda$ is the surface emissivity. This formula assumes a Lambertian (isotropic) distribution of intensity. We thus neglect the beaming (increase of intensity at low phase angles) that results from rugosity and craters. It would be interesting in a future study to include an anisotropic emission to quantify the photometric effect of thermal beaming.

 In practice, $\cos\alpha_j$ is calculated as the scalar product of the direction vectors attached to the center of the planet and pointing towards the observer and the center of the $j$ cell. Only locations visible to the observer ($\cos\alpha_j> 0$) contribute to $\phi_{P,\lambda}$.

The spectral energy distribution of the stellar irradiation is not needed to compute the planetary emission, which is sensitive only to the bolometric luminosity and the Bond albedo. However, it is convenient to express the flux received by the planet (in a given spectral band) in terms of the planet/star contrast ratio. To do this we approximate the stellar radiation by a blackbody at the effective temperature of the star.  When considering a specific stellar system or a comparison with real observations it is more appropriate to use a more realistic spectral density obtained from a model and/or observations, as we did in a separate study dedicated to 55~Cnc~e~\citep{Bolmont_55Cnc}.
\begin{figure*}
\centering
	\includegraphics[width=\figwidthmed]{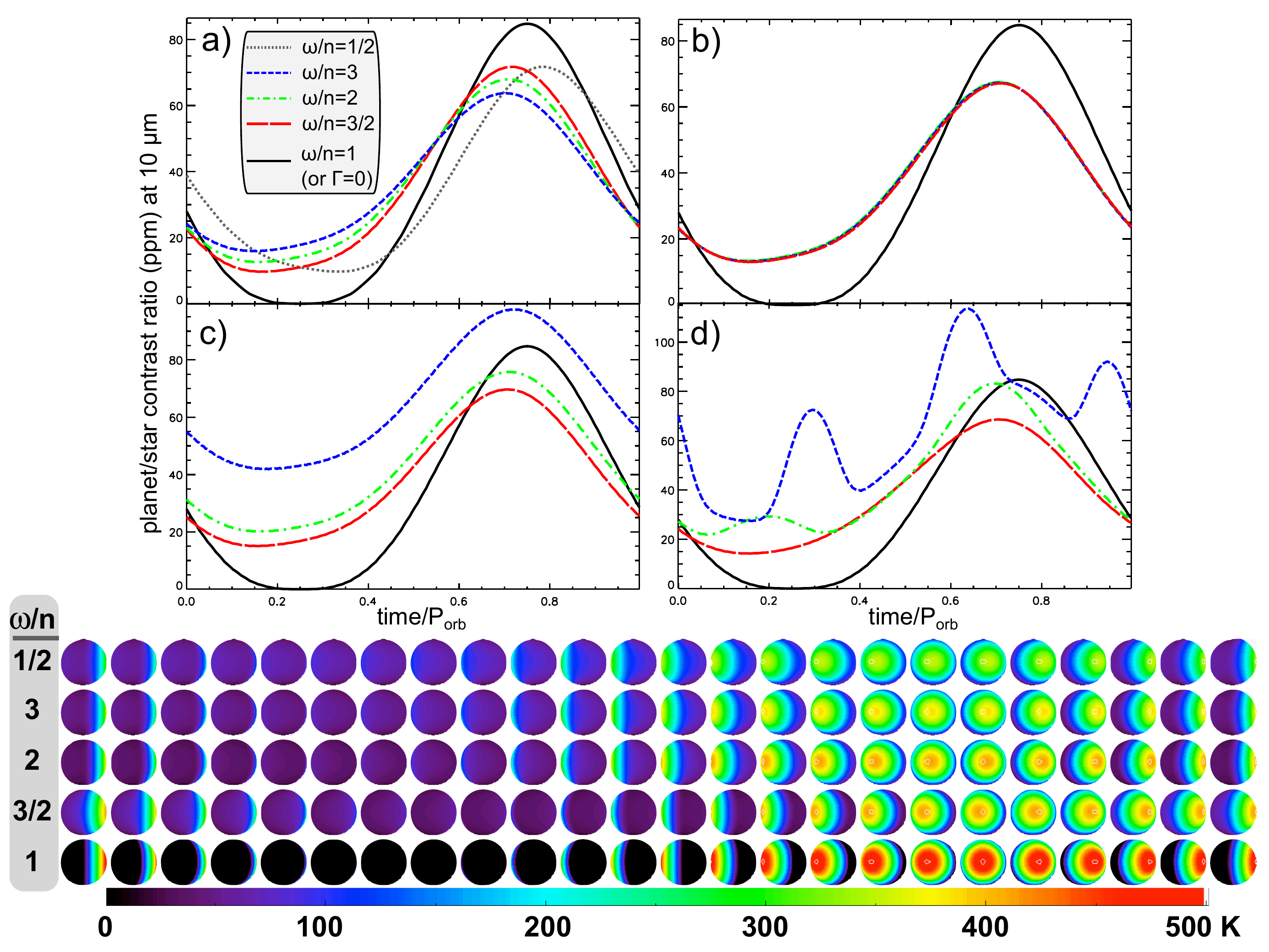}
 	\caption{Light curves and temperatures maps (as seen by the observer) for a circular orbit. \textbf{a)} These light curves are calculated for the default parameters (except for the eccentricity, which is null) and for different rotation periods. Tidal dissipation is not included. Only panel \textbf{a)} includes the case $\omega/n = 1/2$. \textbf{b)} The light curves for $\omega/n = 3/2$ and $3$ are superimposed with that of $\omega/n = 2 $ because their thermal inertia has been modified in order to keep the quantity $\Gamma\sqrt{\omega_\star}$ constant. \textbf{c)} same as \textbf{b)} but with the tidal dissipation included. \textbf{d)} same as \textbf{c)} but the half of the tidal dissipation is released in a localized hot spot. The temperature maps correspond to panel \textbf{a)}. As in all following figures, these maps are displayed for the same orbit and duration as the lightcurves above and are separated by a constant time interval: one $24^{\rm th}$ of the full duration (which can be 1 or 2 orbits depending on the figure). By default, time 0 corresponds to a phase angle of $90^\circ$ (and periastron when $e \ne 0$).}
	\label{fig:circ}%
\end{figure*}

\section{Results} \label{sec:results}
\subsection{Circular orbits} \label{sec:circ}
We first consider a planet on a circular orbit and the effect that a non-synchronized rotation has on the thermal light curve and on tidal heating signatures. In reality, planets that are sufficiently close to their star to exhibit an observable emission (above $10^{-5}$ the stellar flux) are subjected to strong tidal interactions that synchronize very rapidly the rotation with the orbital motion. We considered five planetary rotation periods: $\omega/n=0.5$, $1$, $1.5$, $2$ and $3$. 
These configurations would not realistically be maintained for long given the effects of tidal dissipation.  Rather, any non-synchronous states should actually evolve extremely rapidly towards synchronization, as illustrated in Figure~\ref{fig:synchro}.  Indeed, the planet initially rotating 3 times per orbit slows to 2 rotations per orbit in 10,000 years and is synchronized in less than 100,000~years. It is therefore unlikely that such a configuration would be observed.  Nonetheless, we use the simple case of planets on circular orbits with varying rotation states to illustrate how eccentricity breaks degeneracies and gives rise to a rotational signature.


Figure~\ref{fig:circ} shows the phase curves of for planets on circular orbits with a range of spin states.  All other parameters were set to their default values listed in table~\ref{table:default}. Tidal dissipation is not included. For a circular orbit, the synchronized case is equivalent to a null thermal inertia as all points on the planetary surface receive a constant illumination. In non synchronized cases, the surface thermal inertia has two effects. First, it damps the amplitude of the light curve by lowering the zonal temperature gradient. Second, it induces a phase shift of the light curve, corresponding to a delay of its maximum (westward shift of the hottest point) if the rotation is faster than synchronization, and to an advance (eastward shift of the hottest point) if rotation is slower than synchronization, which is the case for the 1:2 spin-orbit resonance. A projected inclination of $0^{\circ}$ (a face-on orbit) would produce a flat curve (unlike in the case of an eccentric orbit), as would an infinite thermal inertia for any observer position even if the planet is not synchronized. 

Panel~\textbf{b} of Fig.~\ref{fig:circ} illustrates the degeneracy between rotation rate, thermal inertia and albedo.  This is well known for Solar System small bodies: in the absence of internal heat flow, the light curve is indeed controlled by a single thermal parameter proportional to $\Gamma\sqrt{\omega_\star}/T^3_{\rm eq}$ \citep{Spencer1989,Lagerros1996}. By modifying the thermal inertia of the $\omega/n=3/2$ and $3$ cases to match the same $\Gamma\sqrt{\omega_\star}$ as the $\omega/n=2$ case, the same light curves are obtained.

When including tidal dissipation and assuming the same dissipation factor $k_{2,{\rm p}} \Delta t_{\rm p}$ the different rotation rates produce different internal heat fluxes.  The heat flux increases with the departure from the synchronous rotation. Tidal heating increases the planetary flux and lowers the amplitude of the light curve by attenuating the day/night temperature contrast, as shown in panel~\textbf{c} of Fig.~\ref{fig:circ}.  Although the previous degeneracy is broken, inferring the rotation rate requires a priori knowledge of the dissipation. If, however, the tidal dissipation is not released uniformly at the surface but mainly through one or a few large hot spots, like on Io \citep{Veeder_2012_Io}, then the planet behaves like a lighthouse and its light curve exhibits a modulation at the rotation period.  In the phase curves in panel~\textbf{d} of Fig.~\ref{fig:circ}, half of the tidal dissipation is released through one hot spot centered on the equator. This region of increased internal heat flow is modeled with a Gaussian profile and a halfwidth of $45^{\circ}$. Changing the number, location, intensity and size of hot spots would of course affect the amplitude of the rotation modulation.

\subsection{Eccentric orbits}

The degeneracy between thermal inertia and rotation rate is broken for eccentric planets. The combination of the change of insolation during the orbital motion and the rotation of the planet produces a non-homogeneous zonal heating: two meridians do not receive in general the same insolation when they face the star. Due to the surface thermal inertia, the resulting zonal temperature differences produce a photometric variation associated with the rotation that cannot, a priori, be obtained with a different set of rotation rates and thermal inertia. It is easier to understand the phenomenon for a very eccentric orbit. In this case the periastron passage is very brief and produces an intense heating that is limited to a small area of the planet, creating a hot region. After this \textit{periastron branding} the rotating planet shines as an IR lighthouse, producing a rotation modulation that decays as the heated area cools down. At the next periastron passage, it is either another area of the planet that is heated or the same one for some spin-orbit resonances (1:2, 2:1, 3:1, ...). In all cases, the photometric signature is periodic for a uniform planet because the shift between a heated region and the previous one remains constant. \\
\begin{figure*}[p]
	\centering
	\includegraphics[width=\figwidthmed]{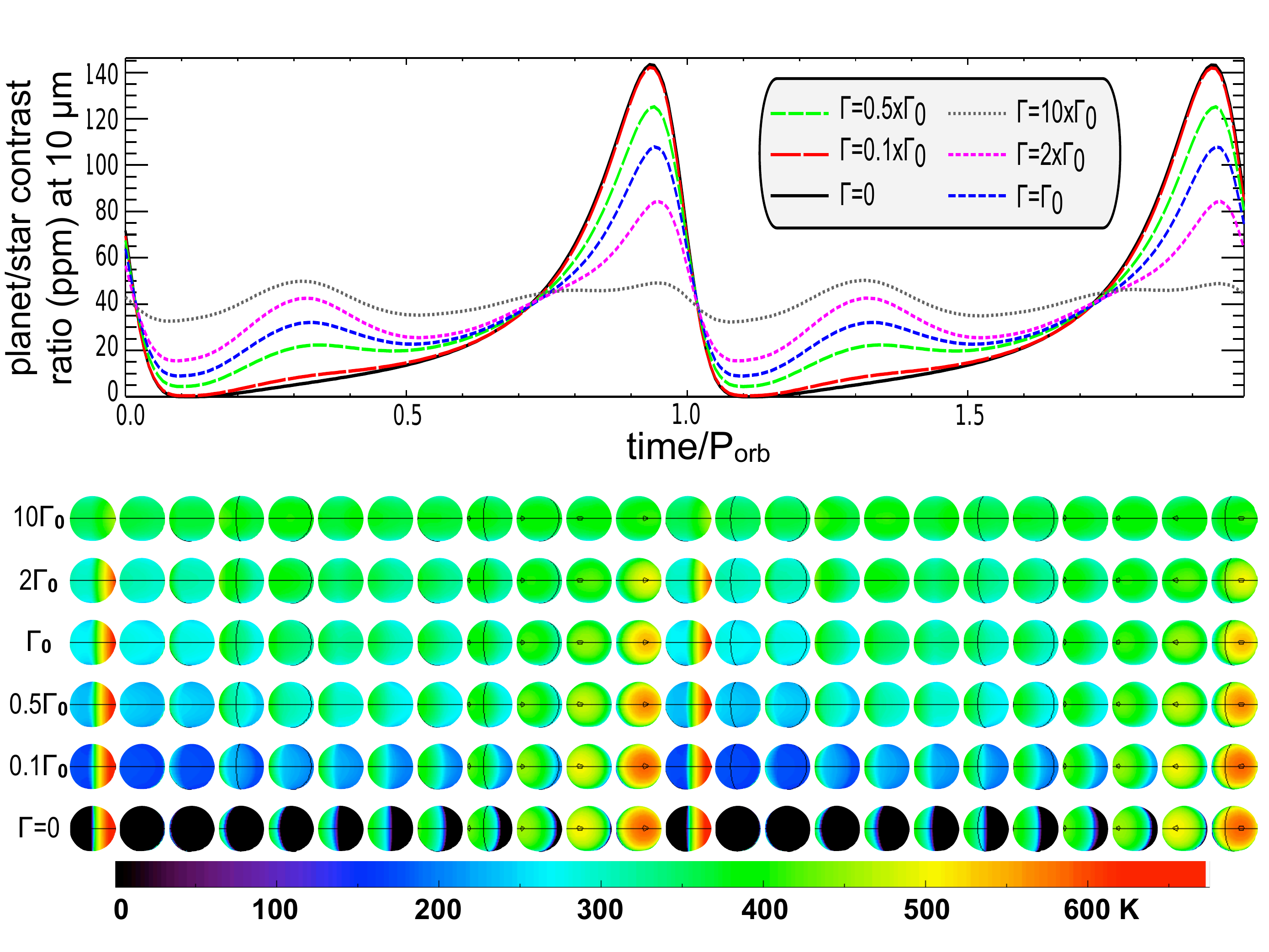}
 	\caption{Effect of the thermal inertia on a pseudo-synchronized planet ($\omega/n=2.38$) with the default parameters of Table~\ref{table:default}. No tidal heat flux is included. The light curves and the surface temperature maps correspond to the same observation geometry and cover the two same orbits. On the temperature maps, the equator and the reference meridian (longitude $0^\circ$ in our coordinate system) are indicated .}
	\label{fig:TI_var}%
	\includegraphics[width=\figwidthmed]{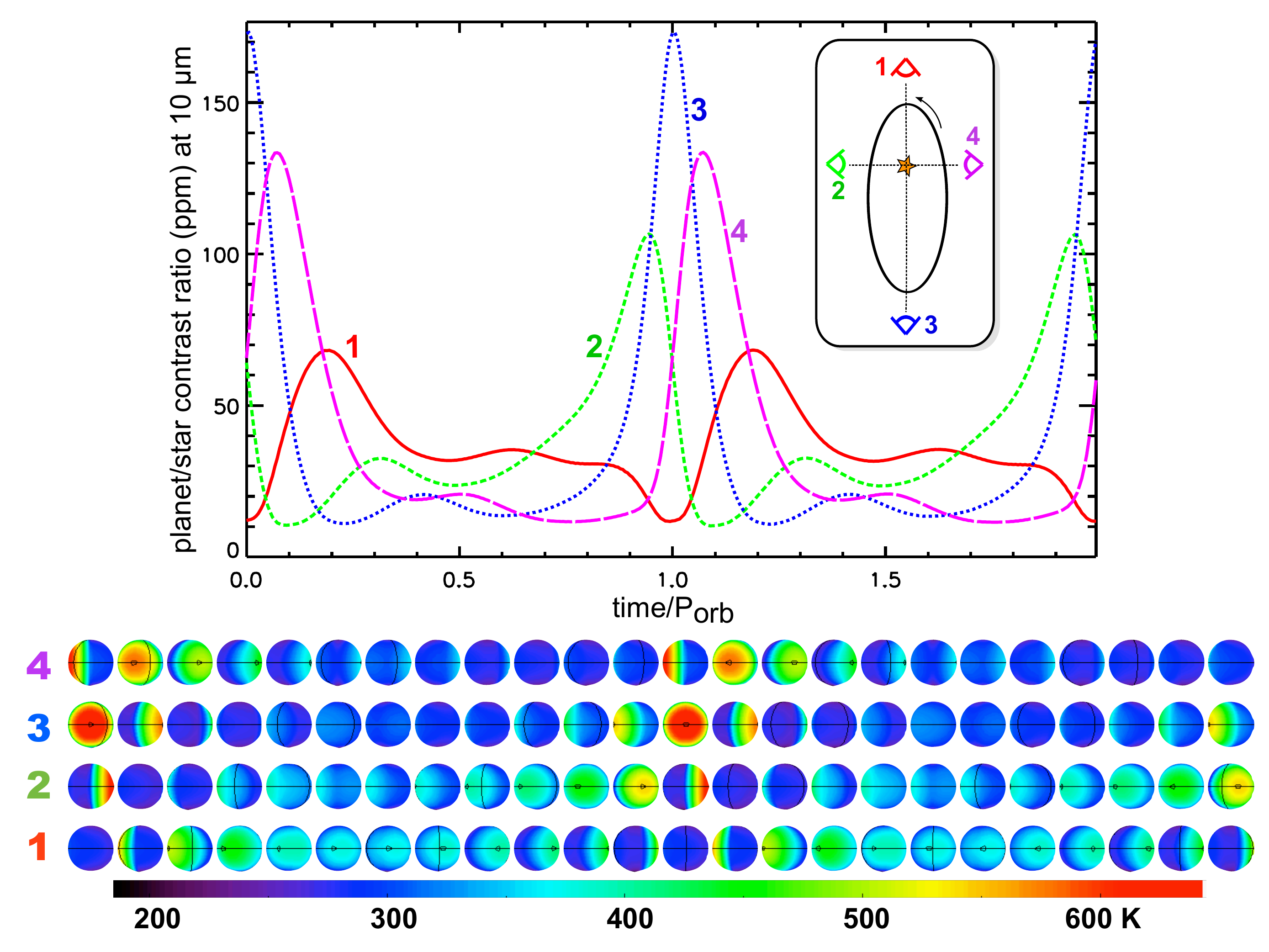}
 	\caption{Same system as above but for $\Gamma_0$ and for different "ecliptic" longitudes of the observer. }
	\label{fig:long_var}%
\end{figure*}
\subsubsection{Influence of the thermal inertia} \label{sec:ti}
The surface thermal inertia must be high enough for the rotation to produce an observable signature. This is why our default thermal inertia is 3000~SI (i.e. Js$^{-1/2}$K$^{-1}$m$^{-2}$). This value is typical of rocks but much higher than that of solar system airless bodies: 50~SI for the Moon, 80~SI for Mercury \citep{Spencer1989}.  For asteroids it varies from a few tens of SI to a few hundreds SI \citep{Delbo2009} and exceptionally reaches 1000~SI \citep{Shepard2008}. Low thermal inertia are due to the presence of a regolith, while bare, solid rocks have a much higher thermal inertia. On the surface of Mars, the thermal inertia varies from about 50~SI in areas covered with fine dust and very little bedrock, to 2500~SI for bedrock~\citep{Putzig2007}. Note that on Mars the thin atmosphere affects the effective thermal inertia of porous or dusty soils, as the heat transport by the gas particle increases the thermal conductivity of the medium compared to the heat transfer in vacuum.

Figure~\ref{fig:TI_var} illustrates the effect of thermal inertia on planetary phase curves, using the default values from Table 2.  Given the expected precision of 10~ppm attainable by EChO and JWST for quiet stars, Fig~\ref{fig:TI_var} shows that 1000~SI is roughly the minimum thermal inertia needed to have a measurable deviation from the lightcurves obtained with a null thermal inertia.  Of course, there are no direct constraints on the (sub)surfaces of exoplanets of a few $\Mearth$; these could be partially made of materials other than regolith such as rocks and metals, solid or melted.  A high surface thermal inertia cannot be ruled out.  A layer of liquid at the surface or in the subsurface can result in a high thermal inertia. For instance, the 25~m average mixed layer at the surface of Earth's oceans has an equivalent thermal inertia of $\sim 25,000$~SI (F. Forget, private communication). A melted (sub)surface is consistent with the high rates of tidal dissipation and resulting internal heat flow associated with eccentric short-period orbit \citep{Barnes2010}. A melted surface implies an atmosphere, which would be inconsistent with our airless assumption. However, vapor pressures of refractory materials are low and the resulting atmospheres could have negligible opacities and negligible influence on the emission and the surface temperature. For instance, a pressure below 1.5~Pa was calculated by \citet{Leger2011} above a substellar magma ocean on CoRoT~7b. The highest value used for thermal inertia in this this graph, $10\times \Gamma_0$, is probably unrealistic as it exceeds that of pure iron but it illustrates the effect of increasing $\Gamma$: for an infinite thermal inertia the surface behaves as if it were exposed to the time-averaged insolation and produces a flat light curve. 

If observations of a given planetary system were to show a significant departure from the $\Gamma=0$ light curve, it would represent an extremely valuable target.  An elevated thermal inertia -- with a minimum value determined by the sensitivity of the measurement (see Fig. 5) -- would provide a direct constraint on the nature of the planetary surface.  In addition, the rotation period can be inferred or at least constrained for planets with large thermal inertia.  This of course requires that the existence of an atmosphere could be discarded by other means; that will be the subject of a future study but is already discussed in \citet{Maurin2012}.  

We do not address here the case of a planetary surface with an inhomogeneous thermal inertia, for instance with large high-inertia spots on an overall low-inertia surface, or vice-versa.
\subsubsection{Influence of the observation geometry}
\begin{figure*}[ph]
	\centering
	\includegraphics[width=\figwidthmed]{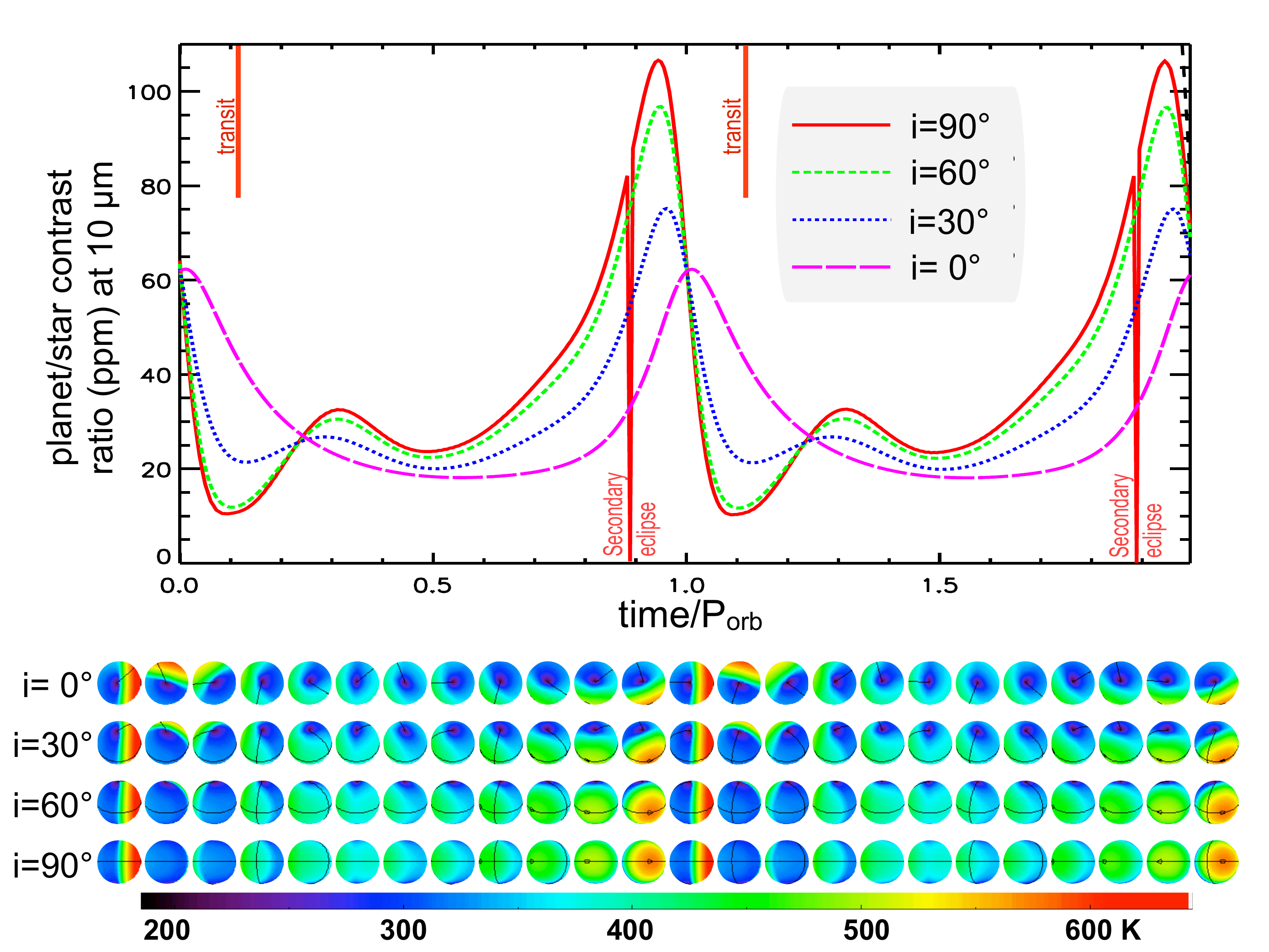}
 	\caption{Influence of the inclination on the light curves and apparent temperature maps. (Default parameters, equilibrium rotation: $\omega/n=2.38$, tidal heat flux not included). For the inclination of $90^\circ$ the secondary eclipses appear on the light curves and the moments of transit are indicated. }
	\label{fig:lat_var}%
 
 	\includegraphics[width=\figwidthmed]{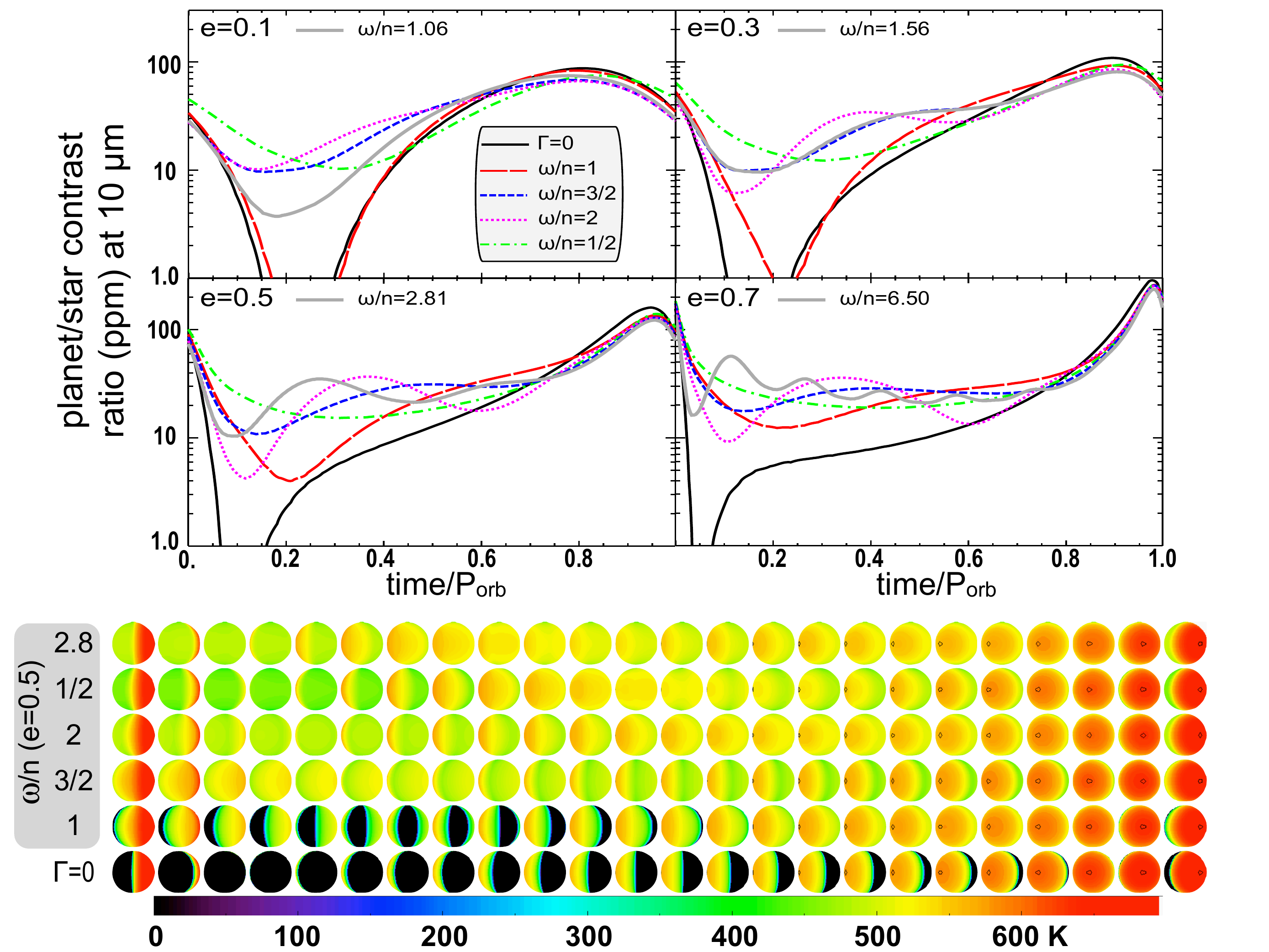}
 	\caption{Top: Light curves for different eccentrics and rotation rates. The solid gray lines indicate the equilibrium rotation (pseudo-synchronization). Bottom: Surface temperature maps for e=0.5. Parameters other than eccentricity are set to default values and tidal heat flux is not included.}
	\label{fig:rot_var}%
\end{figure*}

For a given inclination (or sub-observer latitude), Figure~\ref{fig:long_var} shows that the shape of the light curve depends strongly on the ecliptic\footnote{Here, "ecliptic" refers to the plane of the exoplanet orbit.} longitude of the observer.  It matters whether one observes the day side of the planet near the periastron or the apoastron (even for a null thermal inertia), or whether one observes the quarter that precedes or follows the periastron.  For example, when comparing the light curves seen by observers 1 and 3 in Fig.~\ref{fig:long_var}, the much larger amplitude seen by observer 3 comes from the fact that observer 3 sees the planet's day side during periastron whereas observer 1 does not.  A precise knowledge of the orbital ephemeris from radial velocity or inter-transit duration measurements is therefore required to further analyze the light curves in terms of rotation and thermal inertia. \\

Our default inclination in this study is $90^\circ$ (transit configuration) but the light curve variations are in theory observable for any inclination~\citep[see][for the effect on reflected light curves]{Kane_Gelino_inclination_2011}.  Unlike for circular orbits, a photometric modulation is observable at null inclination due to the variation of the planet reflection and emission with orbital distance. Fig.~\ref{fig:lat_var} shows how the observed thermal light curve changes with the inclination.  The amplitude of the light curve variations is of course the largest for $i=90^\circ$, an inclination that also provides constraints on the radius and the absolute planetary flux from to primary transit and secondary eclipse observations, respectively. However, the $i=90^\circ$ and $i=60^\circ$ light curves differ by less than 10\%. Half of randomly oriented orbits present an inclination between $60$ and $90^\circ$ and similarly strong observable photometric signatures. 

Note that we represented the secondary eclipse (the planet disappearing behind the stellar disk) only in Fig.~\ref{fig:lat_var} although all the other figures are also done for $i=90^\circ$. Our code can produce light curves that include the secondary eclipses but we chose not to show them as we focus on the orbital photometry. Primary transits do not affect the planet's emission but do somewhat increase the planet/star contrast ratio due to the apparent dimming of the star. For our default parameters, however, there is only a 0.3\% increase in contrast ratio.
 \subsubsection{Interplay between eccentricity and rotation}
Figure \ref{fig:rot_var} shows light curves for different eccentricities and rotation rates. Several spin-orbit resonances are considered: 1:2, 1:1 (synchronization), 3:2, 2:1 as well as pseudo-synchronization. 

First, one can notice that, unlike circular orbits, eccentric orbits do not exhibit the same light curve for a null thermal inertia and synchronization. The surface of an eccentric synchronized planet does not receive a constant illumination because of changes in orbital distance and so-called optical librations from the variations of the orbital motion angular velocity.  The position of the substellar point oscillates and the permanently dark area of the planet is smaller than a full hemisphere. This dark region can be seen as an "eye of Sauron" \citep{Tolkien1954} in the temperature map of Fig.~\ref{fig:rot_var} calculated for $\omega/n=1$ and on Fig.~\ref{fig:sauron} . For a null obliquity, this region covers a fraction $s$ of the planetary surface given by the following series expansion
  \begin{align}
 s&=\frac{1}{2} - \frac{1}{\pi} \left( 2\, e + \frac{11}{48} e^3 + \frac{599}{5\,120}e^5 + \frac{
 17\,219}{229\,376}e^7 \right),
 \end{align}
The exact analytical calculation of this fractional area $s$ is given in Appendix~\ref{sec:dark}. For eccentricities larger than $\sim0.724$, all longitudes receive starlight at some position on the orbit. For our default eccentricity of 0.45, only 20.6\% of the surface of a synchronized planet never receives starlight. Note that \citet{selsisHZ2007} also gives a series expansion for $s$, but which is valid only for eccentricities lower than about 0.3.

The effect of the rotation rate on the light curve increases with the eccentricity. For $e=0.7$ and pseudo-synchronization ($\omega/n=6.5$), one can clearly see the effect of the \textit{periastron branding} at time = 0 and the damped rotation modulation after the periastron passages that produces 6 local maxima in the orbital phase curve. 

It is also important to note that, in particular for high eccentricities, the curves obtained for different rotations but the same thermal inertia converge to similar values near the periastron passage. Again, this is easiest to understand for high eccentricities: near periastron, the variation of zenith angle on a point of the surface of the planet is dominated by the angular velocity of the (fast) orbital motion over the angular velocity of the rotation. As a consequence, the light curve close to periastron provides a constraint on the thermal inertia. Closer to apoastron, it is the rotation angular velocity that dominates the modulation of surface irradiation and the light curve is sensitive to both thermal inertia and rotation. Therefore, rotation and thermal inertia should in theory be inferred, at least for a transiting planet with known inclination and radius, and a reference for the absolute stellar flux obtained during the secondary eclipse.\\
\begin{figure*}[hbt]
	\centering
	\includegraphics[width=\figwidthmed]{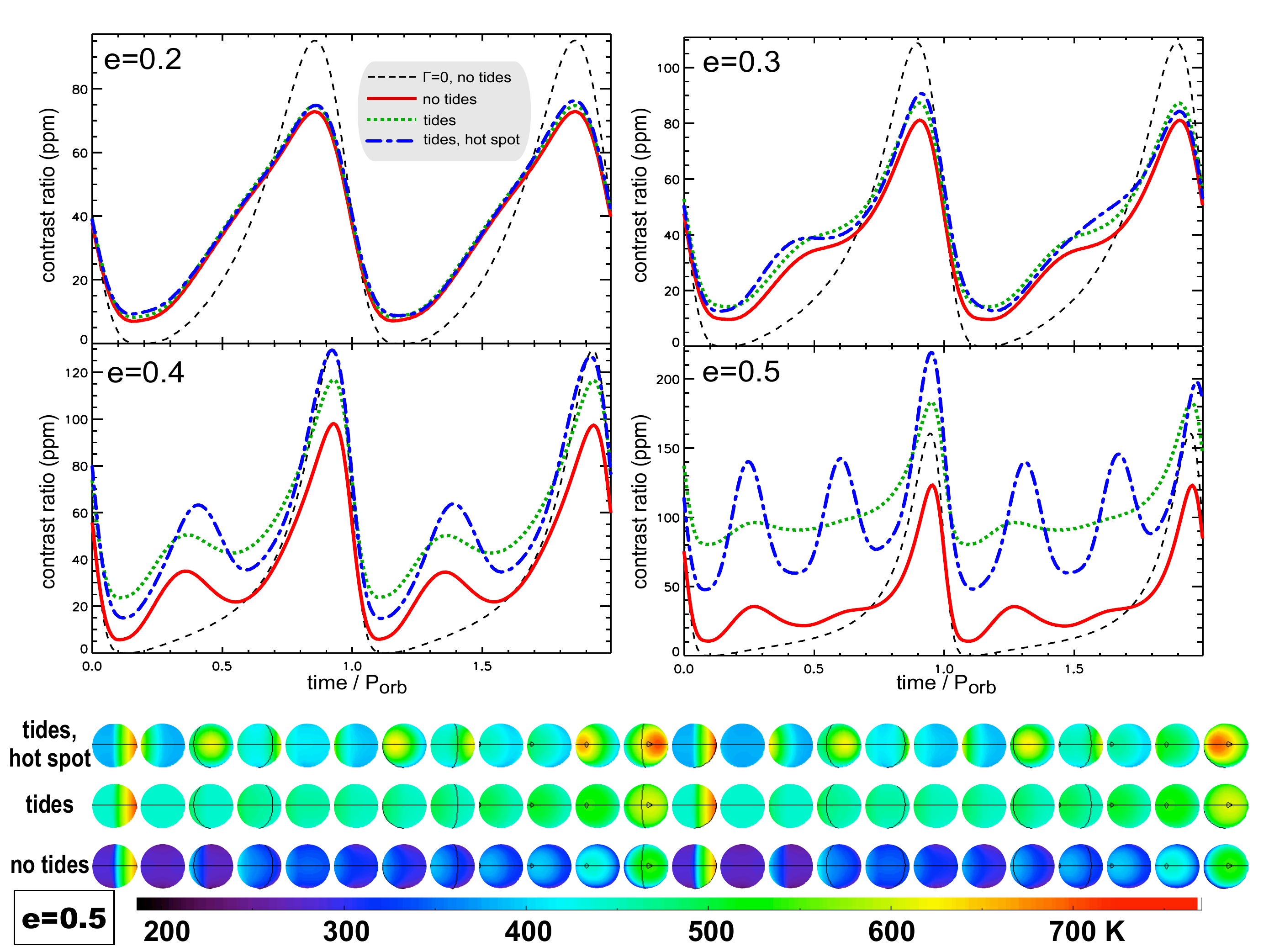}
 	\caption{Effect of tidal heating on pseudo-synchronized planets. All parameters are set to default values except the eccentricity. The dashed line shows the light curve at radiative equilibrium ($\Gamma=0$) and without tidal heating. The solid line is obtained when including the subsurface heat transport but no tidal heating. The dotted line is calculated with heat diffusion and a uniform internal tidal heat flow. To produce the dashed-dotted line, half of the tidal heat is released through a hot spot with a $45^\circ$ radius and centered on the equator (see text). Bottom: Surface temperatures maps corresponding to the light curves calculated for $e=0.5$.}
	\label{fig:tides}%
\end{figure*}

\subsubsection{Influence of the stellar spectral type at constant $T_{\rm eq}$}
\begin{figure}[h]
	\centering
	\includegraphics[width=\figwidthmed]{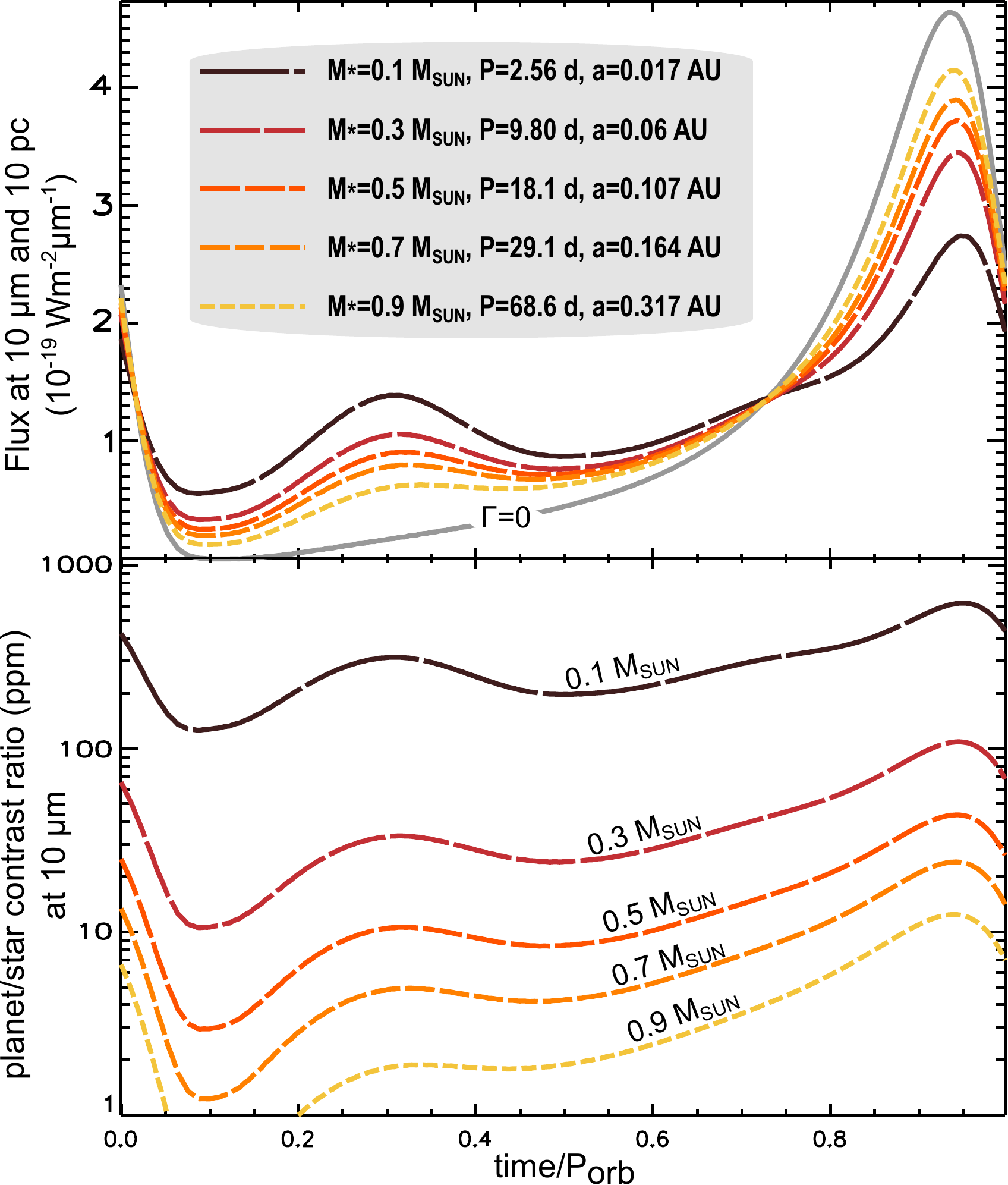}
 	\caption{Lightcurves calculated for a planet receiving an average insolation of 1021 Wm$^{-2}$ ($T_{\rm eq}=366$~K with $A=0.1$) but orbiting different stars. All parameters are set to default values except the stellar mass and orbital period. The flux from the planet received at 10~pc is given in the top panel, while the bottom one shows the planet/star contrast ratio. Corresponding orbital periods $P$ and semi-major axes $a$ are indicated. }
	\label{fig:startype}%
\end{figure}
For a null surface thermal inertia and a given $\omega/n$ ratio the shape and intensity of the thermal light curve does not change with the spectral type if we scale the orbital distance to keep the equilibrium temperature of the planet constant. A non-null thermal inertia makes the thermal light curve sensitive to the orbital period, as illustrated in Fig.~\ref{fig:startype} (top). The irradiation of the surface changes more slowly for a longer orbital period, giving more time for the surface to adapt to the change of illumination. The more massive the star, the longer the orbital period, and the closer the light curve to the case of radiative equilibrium ($\Gamma=0$). 

For a given equilibrium temperature, low-mass host stars provide better candidates to attempt a measurement of the rotation period. The signature of rotation is stronger for short orbital periods.  In addition,  Fig.~\ref{fig:startype} (bottom) shows that the infrared planet/star contrast ratio strongly decreases with increasing stellar mass due to the mass-luminosity relationship.
 
\subsubsection{Influence of tidal dissipation}
We now include tidal heating computed with the constant time-lag model as described in section~\ref{sec:tidal_model}. Note that the terrestrial value we use for the dissipation factor is arbitrary and could differ by orders of magnitude from that of an exoplanet. In particular, the very strong dissipation rates and heat flows considered here would dramatically alter the internal structure and, thus, the dissipation properties of the planet. A more consistent coupling between the dissipation factor and the dissipation rate could be included in the future, provided some assumptions on the composition of the planet.  \\

As in the circular case described in section~\ref{sec:circ}, we assume that the tidal heat is either released uniformly over the planetary surface or half is released uniformly and the other half through a \textit{hot spot}. This hot spot is arbitrarily centered on the equator with a gaussian profile of heat flow and a halfwidth of $45^\circ$. In our model the properties, location and size of the hot spot(s) can be changed but the purpose here is not to present an exhaustive zoology of all the light curves that can be produced by different distributions of surface heat flows. We want to illustrate the fact that a non-uniform heat flow produces a rotation modulation that could be observable for strong tidal heating.\\

Fig.~\ref{fig:tides} shows the resulting light curves with and without the tidal heating for different eccentricities. For the default parameters and $e=0.2$ the combination of rotation and thermal inertia produce some moderate damping of the curve but tides have a negligible effect. For $e=0.3$ the signature of rotation appears as well as a slight increase of the planet brightness. For $e=0.4$ and $0.5$, the tidal heat flux imparts a strong signal on the phase curves. In the case of a uniform release of tidal heat, the signature of rotation is attenuated, in particular at $e=0.5$ as the emission is dominated by the tidal flux. In the case of the hot spot, the rotation modulation is strong and does not decay over an orbit as periastron branding does. Note that in the case of a hot spot, only spin-orbit resonances produce a periodic signal (with a period that can be one or more orbital periods). In Fig.~\ref{fig:tides} the curves are calculated for pseudo-synchronization. This results for $e=0.5$ in $\omega/n=2.84$, and the non-periodicity of the curve can be seen, for instance by looking at successive maxima, which intensity change depending whether the hot spot and periastron heating are superimposed or shifted. \\

Assuming a single strong hot spot obviously maximizes the rotation modulation. Multiple or weaker hot spots would naturally produce a  weaker signature.

\section{Discussions} \label{sec:discussion}
\begin{figure}
	\centering
	\includegraphics[width=\figwidthmed]{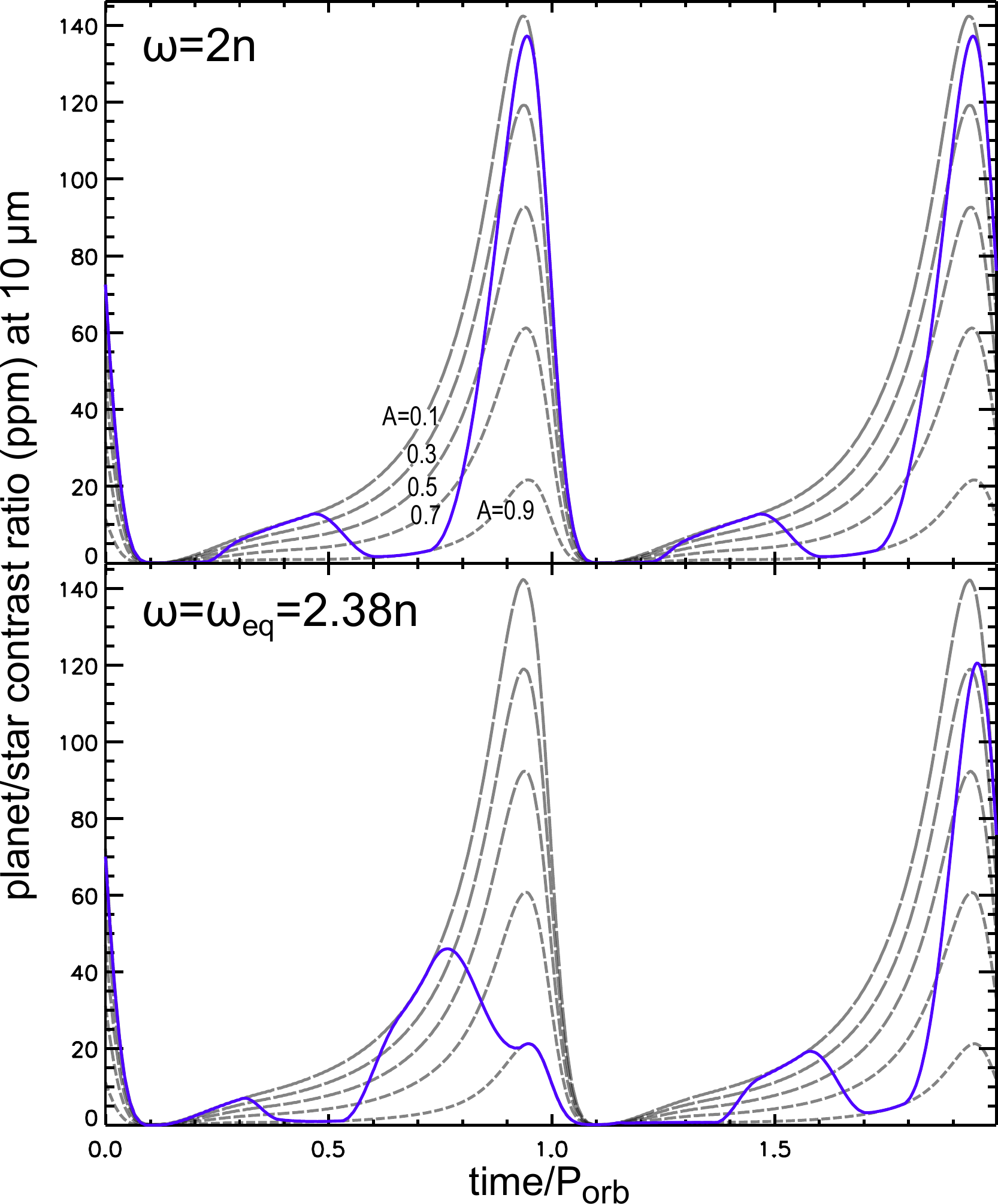}
 	\caption{Light curves for low thermal inertia half-dark half-bright planet. Dashed lines show the light curves for a planet with the default parameters but a low thermal inertia (100~SI) and albedo values between 0.1 and 0.9. The solid curve is obtained if the surface of the planet is divided into two hemispheres (separated by two meridians) with highly different albedos: $A=0.1$ and $0.9$. On the top panel, the system is in a 2:1 spin-orbit resonance. The light curve is periodic but its shape depends on the arbitrary position of the hemispheres. On the bottom panel, the system is in pseudo-synchronization and the signal is no longer periodic.}
	\label{fig:rot_albedo}%
\end{figure}
\subsection{Retrieval of planet parameters}
The retrieval of the rotation period from light curves obtained with future space IR telescopes requires a realistic description of the instrument, sources of noise and stellar variability, and will be the purpose of future studies.  Although such work has not yet been done,  we did perform $\chi^2$ minimizations to test the non-degeneracy between $\omega$ and $\Gamma$. We considered a transiting $2\Rearth$ planet on a 20~day orbit around a 0.3~$M_\odot$ star with an eccentricity of 0.3 (with these parameters the contribution of tidal heating to the thermal emission can safely be neglected). Assuming the radius, inclination and orbital period of the planet are known, which is the case for a transiting planet, the 3 remaining parameters controlling the light curves are $A$, $\omega$ and $\Gamma$. We produced light curves for 2 consecutive orbits for five $2\mu$m-width bands covering the $6-16~\mu$m range on a grid of $A$, $\omega$ and $\Gamma$. We then simulated an observation with EChO \citep{EChO2012} for a set of values: $A_{\rm p}$, $\omega_{\rm p}$ and $\Gamma_{\rm p}$ assuming a distance of 10~pc, a detector throughput of 30\%, and a full coverage of the 2 orbits. Considering only the stellar photon noise, we randomly added a gaussian noise to every exposure ($>24$ per orbit as the result is independent of this number above this value) and produced 500 noisy observations. We computed the individual $\chi^2$ between each noisy light curve and those from the grid and produced a 3D map of the mean $\chi^2$ in the $A$-$\omega$-$\Gamma$ space. We then identified the n$-\sigma$ contours around the actual $A_{\rm p}$, $\omega_{\rm p}$ and $\Gamma_{\rm p}$ values. 

The values we set for the albedo and rotation period are $A_{\rm p}=0.3$ and $\omega_{\rm p}/n=2$ and we considered two values for the thermal inertia: $\Gamma=1500~$SI ($\Gamma_0/2$) and $\Gamma=3000~$SI ($\Gamma_0$). As noted in section~\ref{sec:ti} the specific signature of rotation increases with the thermal inertia and we thus expect a better retrieval with the highest value of $\Gamma$. With $\Gamma=1500$~SI we found the following $2\sigma$ errors: $\omega/n=2^{+0.25}_{-0.1}$, $\Gamma=1500^{+1000}_{-600}$~SI and $A=0.3\pm0.05$. With $\Gamma=3000$~SI, the error decreases as expected: $\omega/n=2^{+0.15}_{-0.05}$, $\Gamma=3000\pm800$~SI and $A=0.3\pm0.05$ (error on the albedo is unchanged). Again, these results are more a practical confirmation that the effects of rotation and thermal inertia, which are indistinguishable for a circular orbit, are broken apart by eccentricity, rather than a robust prediction of what can be expected from real observations. It is however promising that the spin might be constrained even with EChO (a $1.1$~m telescope) while JWST will be able to obtain a much higher SNR with an collecting area 25 times larger. For out-of-transit exoplanets, the radius and the inclination would also have to be constrained and the retrieval of the five parameters has not been tested yet. 

\subsection{Rotation signature at low thermal inertia from albedo spots}
As shown in section~\ref{sec:ti}, a large thermal inertia is needed to impart a measurable signature of planetary rotation on the phase curve. This is true for a planetary surface with uniform properties but not necessarily for a planet with a very heterogeneous albedo. An extreme case is illustrated in Fig~\ref{fig:rot_albedo}.  This figure shows that a non-synchronized planet with low surface thermal inertia but with dark and bright hemispheres (separated by two meridians) produces a rotation modulation that superimposes with the orbital response. A more realistic albedo surface distribution with less contrast and smaller patterns would, of course, result in a weak modulation with an amplitude unlikely above 10~ppm.
\subsection{Anisotropic and spectral radiative properties of the surface}
\begin{figure*}
	\centering
	\includegraphics[width=\figwidthmed]{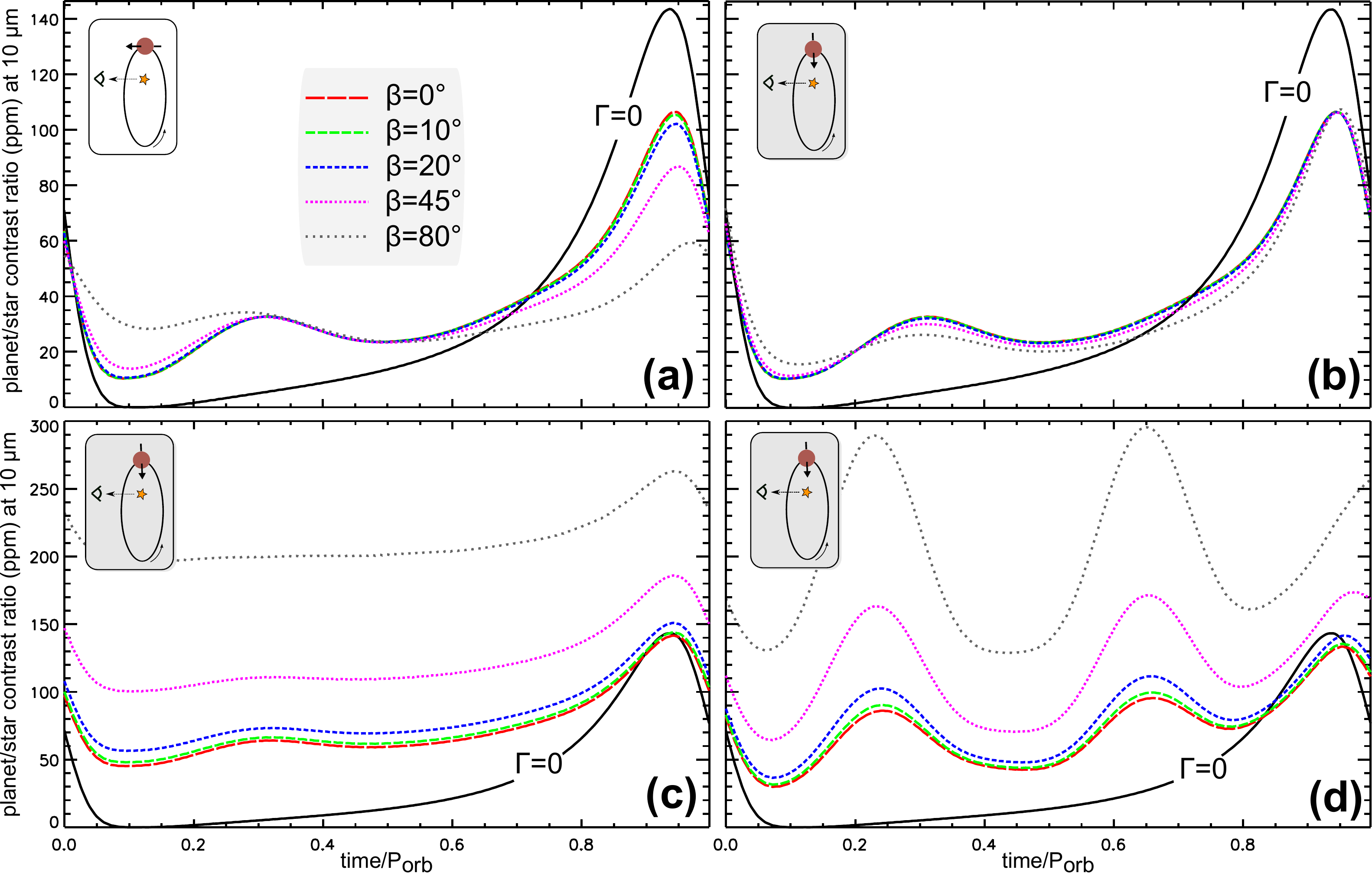}
 	\caption{Influence of the obliquity $\beta$. Parameters are set to their default values and $\omega=\omega_{\rm eq}=2.38 n$. Panels \textbf{(a)} and \textbf{(b)}: light curves for different obliquities and two orientations of the spin axis, without tidal heating. Panels \textbf{(c)} and \textbf{(d)}: Same as \textbf{(b)} but with tidal heating included, uniformly in \textbf{(c)} and with half of the tidal dissipation released through one hot spot (gaussian, half width: $45^\circ$) centered on the equator in \textbf{(d)}.}
	\label{fig:obl}%
\end{figure*}
Our model assumes an emissivity of 1 independent of the wavelength. In reality, planetary surfaces are not blackbodies and their emissivity (although usually close to unity) vary with wavelength, depending on the surface material. There is for instance a well known absorption band around $10~\mu$m for silicates. These spectral properties are used to determine the composition of planetary surfaces in the solar system. The emissivity of Solar System regoliths have been studied in details \citep[e.g][]{Lim2005} but regolith have a low thermal inertia and would not exhibit a measurable signature of rotation. The emissity of high thermal inertia refractory surfaces (bedrocks, magma) is still to be studied. \citet{Hu2012} produced synthetic spectra of planetary surfaces of various compositions that could be expected for highly irradiated exoplanets, but significant emissity spectral features are associated with dusty/pourous the surface \citep{Vernazza2012}. Assuming that a high thermal inertia is inconsistent with a dusty/porous surface, the spectral signatures of high-inertia surfaces is not expected to exhibit strong spectral features.
 
Note that the spectral dependency of the albedo is not important as long as we do not include the reflected light, as we are only sensitive to the surface albedo averaged over the wavelengths of the insolation. 

The surface albedo may depend on the incidence angle, as it does on the Moon and generally on low albedo regoliths. This is why the full Moon appears flatter than a Lambertian sphere. We performed some tests comparing light curves with a Lommel-Seeliger bidirectional reflectance distribution function (BRDF) for the reflected light \citep{Fairbairn2005} and with an equivalent constant Bond albedo. We found negligible effect on the thermal light curve due to the fact that areas under grazing illumination have a low temperature and do not contribute significantly to the disk-integrated emission. The effect on the temperature is very small and the latitude-dependence of temperature on the starlit hemisphere of the Moon (for which we can neglect the very low thermal inertia) follows fairly well a $T_{SS}\mu^{1/4}$ law \citep{Lawson2000}, where $T_{SS}$ is the subsolar point.\\

The thermal emission of Solar System planetary surfaces is known to be slightly anisotropic, with a beaming toward the direction of illumination (similar to the opposition surge known for reflected light). This is due to macroscopic rugosity and craters. For low zenith angles, a crater floor is not affected by the shadows of its edges but, on the contrary, the crater floor receives the thermal emission from the edges in addition to the stellar illumination, resulting in a higher brightness temperature for an observer that is roughly in the direction of the star. The effect of rugosity can be understood if the surface is modeled as a layer of spheres. Each illuminated sphere has as hot substellar point (assuming a low thermal inertia) and emits more in the direction of the illumination. As the opposition effect, this beaming is the strongest for very low phase angles. For non transiting exoplanets and transiting exoplanets, the phase angle for a distant observer cannot be lower than the angular radius of the star, which limits the effect of the beaming. This is in particularly true for very close-in exoplanets for which the angular stellar radius can be very high ($\sim 15^\circ$ for Kepler~10~b and 55~Cnc~e). New thermal modeling of asteroids shows that the beaming effect is not only important at low phase angles, and that the most important effect is coming not from the sub-solar point but from the rough surface elements near the terminator \citep{Rozitis2011}. While we neglect these effects here, it might be important to include them in a forthcoming study.

Taking into account spectral and anisotropic effects of realistic planetary surfaces can be done by computing BRDFs with Hapke functions and associated parameters \citep{Hapke1981}. This may be done in the future, at least to assess the impact on the retrieval of the rotation period but it would imply adding several parameters, as unknown as the nature of exoplanetary surfaces. In addition, Hapke functions have been developed to reproduce the BRDF of regolith and may not be adapted for different types of surfaces that could be found on exoplanets.   
\subsection{Obliquity}
To this point we have assumed that obliquity tides have damped any initial obliquity to zero. For isolated planets with a planet/star IR contrast ratio above 10~ppm, obliquity damping time is indeed very efficient: for our default parameters, an initial obliquity of $80^\circ$ decreases to $5^\circ$ in just $20,000$~yrs. Dissipation factors must be 5 orders of magnitude lower than Earth's to preserve a high initial obliquity for more than 1 Gyr, although this situation cannot be ruled out as uncertainties on the dissipation span many orders of magnitude.  In the presence of planetary companions, however, the planet can be captured into a Cassini state, a resonance between spin precession and orbital precession, with any non-zero obliquity \citep{Colombo1966,Peale1969}. Our model is designed to include an obliquity but this adds to the parameters list the two angles that define the orientation of the spin axis: the obliquity and, for instance, the periastron longitude (which is the angle between the spring equinox and the periastron passage). 

Figure~\ref{fig:obl} shows a limited exploration to test the effect of obliquity on the phase curve.  Without tidal dissipation, the effect on the light curve is small ($<1\%$) if the obliquity is smaller than $10^\circ$ (panels \textbf{(a)} and \textbf{(b)} of Fig.~\ref{fig:obl}).  The viewing and orbital geometries play an important role and can introduce new degeneracies, in particular between thermal inertia and obliquity.  Indeed, in panel \textbf{(a)} of Fig.~\ref{fig:obl} the obliquity produces, for this particular observation geometry, a damping and phase shift of the phase curve that mimic a thermal inertia.\\

Panel graph \textbf{(c)} of Fig.~\ref{fig:obl} shows that, when tidal heating is included with our default dissipation factor,  the planet emission increases with the obliquity. However, unless the dissipation can be inferred by independent measurement, this cannot be used directly to constrain the obliquity. Panel graph \textbf{(d)} of the same figure shows the thermal light curve obtained when 50\% of the tidal heating is released through a hot spot (with arbitrary profile, position and size). Cases like \textbf{(c)} and \textbf{(d)}, in which both the obliquity and the tidal dissipation are high, should not last for long and are unlikely to be observed: the high dissipation rate would either have eroded the initial obliquity or prevented/swept away possible Cassini states \citep{Fabrycky2007}.

\section{Conclusions} \label{sec:conclusion}
In this paper we studied the effect of a planet's rotation on its infrared photometric lightcurve.  We also included the effect of tidal dissipation within the planet, the strength of which is a strong function of the planet's orbital eccentricity (see Eqn~\ref{equ:dissip}). We generated a wide range of synthetic IR phasecurves to determine the most important parameters.  \\

For an airless planet on a circular orbit, the effect of surface thermal inertia and rotation on the thermal lightcurve cannot be distinguished. This degeneracy is broken for an eccentric orbit by an effect we call \textit{periastron branding} -- the peak of heating at periastron produces a longitudinal gradient of temperature, which in turns result in a rotation modulation -- allowing for the rotation and the surface thermal inertia to be measured or, at least, constrained.\\

For an eccentric planet captured into the 1:1 spin-orbit resonance we provide an analytical expression as well as a series expansion for the fractional area that never receives starlight as a function of eccentricity (Appendix A below).\\

The higher the eccentrity and surface thermal inertia, the better the constraint on the rotation period. Below a value of about 1000~SI, rotation and thermal inertia have a negligible effect on the thermal lightcurve. Constraining these parameters requires a \textit{Super Mercury} with a \textit{super} thermal inertia. This implies solid or melted, non-porous surface different from the regoliths found on airless bodies in the Solar System.\\

Tidal heating can affect the radiative budget of the planet and its thermal lightcurve. If released homogeneously over the surface of the planet, tidal heat increases the overall infrared emission and damps the thermal lightcurve variations associated with the insolation and rotation. If there is a longitudinal asymmetry in the surface tidal heat flux (like on Io) then the rotation signature can be enhanced.  With hot spot tidal heating, a planet behaves as an IR lighthouse regardless of the thermal inertia. Albedo or thermal inertia spots can produce a similar "lighthouse" effect. \\

A photometric precision of 10~ppm, stable over at least two orbital periods, is needed to retrieve the rotation period, in the best case scenario. This is compatible with the planned characteristics of EChO \citep{EChO2012}.  Assuming photon-noise limited observations and a perfectly stable star, preliminary retrieval calculations are promising for EChO and JWST.  However, we note that these calculations were made for synthetic eccentric and transiting exoplanets whereas the known candidates with significant eccentricities are out-of-transit (see Table 1). Stellar variability and instrument stability \citep[e.g.]{Crossfield2012} are expected to be the main obstacle.\\

Measuring the rotation period of tidally-evolved eccentric planets would be extremely important to test tidal models in conditions that cannot be studied in the Solar System. The rotation period can have dramatic consequences on climate and habitability and the first potentially habitable planets discovered, GJ581d \citep{selsisHZ2007,Makarov_GJ581_2012,Wordsworth2011} and GJ667Cc \citep{Delfosse2012} are both eccentric and subjected to strong tides. The study of \textit{Super Mercuries} thermal light curvescould provide a better understanding of tidal interactions and, indirectly, on habitability.

\begin{acknowledgements}
      FS acknowledges support from the European Research Council
(Starting Grant 209622: E$_3$ARTHs). JL acknowledges support from the DIM ACAV. FH and SNR abide. We thank the referee for helpful comments.
\end{acknowledgements}
\begin{appendix}
\section{Analytical calculation of the permanently dark area for synchronous planets on eccentric orbits} \label{sec:dark}
 \begin{figure}[htb]
	\centering
	\includegraphics[width=\linewidth]{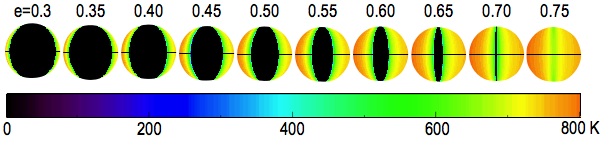}
 	\caption{The \textit{eye of Sauron}. Temperature maps showing the shrinking of the permanently dark area with increasing eccentricities for a synchronously rotating planet. Except for $e$, default parameters are used.} 
	\label{fig:sauron}%
\end{figure}
For a synchronous planet with a null obliquity on an eccentric orbit, the fraction of the surface that receives starlight at some point of the orbit  is greater than 1/2 because the varying orbital motion will cause a libration of the star as seen from the planet. In other words, the longitude of the substellar point
\begin{equation}
 \textrm{lon}_{\star}(t) = \nu(t) - \omega t + C 
 \end{equation}
 varies with time (see Sect.\,2.1). The amplitude of the libration is given by extrema of $\textrm{lon}_{\star}(t)$ that we calculate hereafter. 
 
 Because $\textrm{lon}_{\star}(t)$ is a smooth function, we search for the values of $t$ that cancel its derivative. However, we cannot use the time variable, which requires to solve Kepler's equation. The idea is thus to use the true anomaly as our variable (the change of variable is smooth) and to search for $\nu_\mathrm{ext}$ such that 
 \begin{equation}\label{tosolve}
 \frac{\mathrm{d}\,\textrm{lon}_{\star}}{\mathrm{d} \nu}(\nu_\mathrm{ext}) = 1 -  \frac{\mathrm{d}}{\mathrm{d} \nu}(\omega t)  =1 -  \frac{\omega }{\dot{\nu}}  =0,
 \end{equation}
 where the $\dot{}$ operator stands for the temporal derivative.
For a Keplerian orbit, 
 \begin{equation}\label{nudot}
\dot{\nu} (\nu) = n \frac{\left(1+e\cos\nu\right)^2}{\left(1-e^2\right)^{3/2}},
 \end{equation}
 where $n$ is the orbital mean motion.
Substituting Eq.\,(\ref{nudot}) in Eq.\,(\ref{tosolve}), recognizing that $\omega=n$ for a synchronous rotation, and keeping only the real solutions, we find 
\begin{equation}\label{nuext}
\cos \nu_\mathrm{ext} = \frac{-1+\left(1-e^2\right)^{3/4}}{e},
 \end{equation}
which defines the locations ($\nu_\mathrm{ext}\equiv\nu_+$ and $\nu_-$) of our two extrema over one orbit. To find the values of the extrema themselves, we have to integrate 
\begin{equation}
\textrm{lon}_{\star}(\nu_\mathrm{ext} ) = \int_{t_0} ^{t(\nu=\nu_\mathrm{ext} )} \left(\dot{\nu}-\omega\right) \mathrm{d} t =\int_{\nu_0} ^{\nu_\mathrm{ext} } 1- \frac{\omega}{\dot{\nu}} \mathrm{d} \nu\,,
 \end{equation}
where $\nu_0$ is an arbitrary choice for the reference of the true anomaly that 
 changes $C$ but not the physical results. Again using Eq.\,(\ref{nudot}), we find that 
\begin{align}\label{nut}
\textrm{lon}_{\star}(\nu_\mathrm{ext} ) = \nu_\mathrm{ext}&+C+\frac{e\sqrt{1-e^2} \sin \nu_\mathrm{ext}}{1+e \cos \nu_\mathrm{ext}}\nonumber\\
&-2\, \arctan\left( \frac{\left(1-e\right)\tan\left(\frac{\nu_\mathrm{ext}}{2}\right)}{\sqrt{1-e^2}}\right).
 \end{align}
 
 Finally, the fraction of the surface that never receives daylight is given by
 \begin{align}
 s&=\frac{1}{2}-\frac{\textrm{lon}_{\star}(\nu_+ )-\textrm{lon}_{\star}(\nu_- )}{2\,\pi} ,
 \end{align}
 which can be computed using Eq.\,(\ref{nuext}) and Eq.\,(\ref{nut})
 When this fraction becomes negative (around $e_0\approx0.724$), it just means that the anti stellar point at the periastron passage receives light once between periastron and apastron, and once between apastron and periastron. Over the range of eccentricities where $s$ is meaningful ($\left[0,e_0\right]$), the first terms of  expansion of this function,
  \begin{align}
 s&=\frac{1}{2} - \frac{1}{\pi} \left( 2\, e + \frac{11}{48} e^3 + \frac{599}{5\,120}e^5 + \frac{
 17\,219}{229\,376}e^7 \right),
 \end{align}
 reproduce the latter with an absolute error below 0.002.
 
This calculation assumes a point-like star and should be corrected to account for the angular extension of the star, which reduces the value of $s$ and the critical eccentricty at which this area disappears.
\end{appendix}
\bibliographystyle{aa}
\bibliography{bib_rotides2.bib}

\end{document}